\documentclass[aps,showpacs,preprint,footinbib,preprintnumbers]{revtex4}
\usepackage{amssymb}
\usepackage{amsmath}
\usepackage{amsfonts}
\usepackage{bm}
\usepackage{here}
\usepackage{graphicx}

\newcommand{\be}{\begin{equation}}
\newcommand{\ee}{\end{equation}}
\newcommand{\bea}{\begin{eqnarray}}
\newcommand{\eea}{\end{eqnarray}}
\newcommand{\del}{\partial}

\newcommand{\nonum}{\nonumber}

\begin{document}

\title{Kaon-Nucleon scattering states and potentials in the Skyrme model}

\author{Takashi Ezoe$^1$ and Atsushi Hosaka$^{1,2}$}
\affiliation{$^{1}$Research Center for Nuclear Physics, Osaka University, Ibaraki, 567-0048, Japan}
\affiliation{$^{2}$Advanced Science Research Center, Japan Atomic Energy Agency, Tokai, Ibaraki, 319-1195 Japan}

\date{\today}
\begin{abstract}
We study the~(anti)kaon nucleon interaction in the Skyrme model. 
The kaon field is introduced as a fluctuation around the rotating Skyrmion for the nucleon.  
As an extension of our previous work, we study scattering states and examine phase shifts in various kaon-nucleon channels.  
Then we study the interaction, where we find that it consists of central and spin-orbit components for isospin channels, $I = 0, 1$, with energy dependence and nonlocality.  
The interaction is then fitted to a Shr\"{o}dinger equivalent local potential for s- and p-waves.  

\end{abstract}
\maketitle
\section{Introduction}
The kaon-nucleon system is one of interesting systems in hadron physics.
It is considered that the anti-kaon and nucleon~$\left( \bar{K} N \right)$ interaction is strongly attractive.
Based on the properties of the $\bar{K} N$ strong attraction, a lot of discussions for the $\bar{K} N$ systems have been done.
One example is the $\Lambda \left(1405 \right)$ resonance known as a candidate of the $\bar{K} N$ quasi-bound state~\cite{Dalitz 1, Dalitz 2}
whose properties can not be explained easily by a simple quark model~\cite{Isgur}.
Another example is the kaonic nucleus where the anti-kaon is bound to a nucleus by a strong attraction between them.
It is expected that, because of the strong attraction, the structure of the kaonic nucleus is largely modified from normal nuclei~\cite{Yamazaki 1, Yamazaki 2}.
In such discussions, kaon-nucleon interaction is obviously the most important input.  

In this article, we first discuss the phase shift for kaon-nucleon scattering states by a modified bound state approach proposed in the previous work~\cite{ezoe-hosaka}.
Our approach is based on the bound state approach which is proposed by Callan and Klebanov~\cite{Callan-Klebanov 1, Callan-Klebanov 2}.
In the original approach, the kaon is introduced as a fluctuation around the hedgehog configuration, and then the kaon-hedgehog system is collectively quantized as hyperons. 
On the other hand, in our approach, we first generate the nucleon by quantizing the hedgehog soliton, and then introduce the kaon fluctuation around the physical nucleon.
The difference of the Callan-Klebanov and our approaches is the ordering of projection and variation.
The Callan-Klebanov approach corresponds to the projection after variation, while ours to the variation after projection.
In the previous paper, we have investigated $\bar{K} N$ bound states.
As a result, we found one bound state for the $\bar{K} N \left( J^P = 1/2^-, I = 0 \right)$ channel with a binding energy of order ten MeV corresponding to $\Lambda(1405)$.  

Secondly, we derive a Schr\"{o}dinger equivalent local potential for the kaon and nucleon.  
The resulting potential is fitted by Gaussian type functions which is convenient for the study of few-body nuclear systems with the anti-kaon.
In general, the kaon-nucleon potential has four components; 
the isospin independent and dependent central terms, and spin-orbit terms~(LS terms).
These complete all possible components for the pseudo-scalar and iso-scalar kaon and the spinor and iso-spinor nucleon.
Furthermore, the interaction is energy dependent and nonlocal.  

We organize the paper as follows.
In the next section, we briefly review our approach which we have constructed in the previous work. 
In Sec.~\ref{sec: scattering state}, we discuss phase shifts for kaon nucleon scattering states with lower kaon partial waves.
In Sec.~\ref{sec: potential}, we derive various components of the potential and perform fitting to Gaussian type functions.  
Then we discuss scaling properties of the potential associated with the scaling properties of soliton solutions.
In the end, we summarize the present work and discuss further studies.

\section{Formalism}
\label{sec: formalism}
In this section, we review our modified bound state approach.
Detailed discussions have been done in Ref.~\cite{ezoe-hosaka}.
Let us start with the following Lagrangian for the SU(3)-valued field $U = U \left( r \right)$
\bea
	L = \displaystyle{\frac{1}{16}} F_{\pi}^2 \mathrm{tr} \left( \del_{\mu}U \del^{\mu}U^{\dag} \right)
		+ \displaystyle{\frac{1}{32e^2}} \mathrm{tr}\left[ \left( \del_{\mu}U\right)U^{\dag}, \left( \del_{\nu}U\right) U^{\dag} \right]^2
		+ L_{SB}
		+ L_{WZ},
\label{eq: lagrangian}
\eea
where the first and second terms are the Skyrme Lagrangians~\cite{Skyrme model 1, Skyrme model 2, Skyrme model 3}
and
the third term is the symmetry breaking term due to finite masses of the SU(3) pseudo-scalar mesons~\cite{SU(3) Skyrme model 1, SU(3) Skyrme model 2}
\bea
	L_{SB} = \displaystyle{\frac{1}{48}} F_{\pi}^2 \left( m_{\pi}^2 + 2 m_K^2 \right) \mathrm{tr} \left( U + U^{\dag} -2 \right)
			+ \displaystyle{\frac{\sqrt{3}}{24}} \left( m_{\pi}^2 - m_K^2 \right) \mathrm{tr} \left[ \lambda_8 \left( U + U^{\dag} \right) \right].
\eea
In this paper, we treat the pion as a massless particle while the kaon as massive one.
We call these three terms in Eq.~(\ref{eq: lagrangian}) as normal Lagrangians in this paper.
The last term in Eq.~(\ref{eq: lagrangian}) is the contribution of the chiral anomaly called the Wess-Zumino term given by~\cite{WZ term 1, WZ term 2, WZ term 3}
\bea
	L_{WZ} = - \displaystyle{\frac{i N_c}{240 \pi^2}} \int d^5 x \  \varepsilon^{\mu \nu \alpha \beta \gamma}
				\mathrm{tr} \left[ \left( U^{\dag} \del_{\mu} U \right)
							\left( U^{\dag} \del_{\nu} U \right) 
							\left( U^{\dag} \del_{\alpha} U \right) 
							\left( U^{\dag} \del_{\beta} U \right) 
							\left( U^{\dag} \del_{\gamma} U \right) \right],
\eea
where $N_c$ is the number of colors, $N_c = 3$.

The Lagrangian Eq.~(\ref{eq: lagrangian}) contains three parameters; 
the pion decay constant, $F_{\pi}$,
the Skyrme parameter, $e$,
and the mass of the kaon, $m_K$.
Here, we keep $m_K$ at the experimental value, 495 MeV, and we consider three parameter sets for $F_{\pi}$ and $e$.
We will show them in the next section.

To study the interaction of the kaon with the physical nucleon, we introduce the ansatz, 
\bea
	U = A \left( t \right) \sqrt{U_{\pi}} A^{\dag} \left( t \right) U_K A \left( t \right) \sqrt{U_{\pi}} A^{\dag} \left( t \right),
\label{ans: our ansatz}
\eea
where $A \left( t \right)$ is an isospin rotation matrix,  $U_{\pi}$ is the Hedgehog pion field with the soliton profile function, $F \left( r \right)$,
\bea
	U_{\pi} 
		= \begin{pmatrix}
			\xi^2	& 0 \\
			0      	& 1
		\end{pmatrix}
	, \ \ \ \ \xi^2  =  \exp \left[ i F \left( r \right) \bm{\tau} \cdot \hat{r} \right],
\label{eq:def of hedgehog soliton}
\eea
and
\bea
	U_K = \exp \left[ \displaystyle{\frac{2 \sqrt{2} i}{F_{\pi}}}
			\begin{pmatrix}
				0             & K \\
				K^{\dag} & 0
			\end{pmatrix}
				\right]
	, \ \ \ \ K = \begin{pmatrix}
				K^{+} \\
				K^{0}
			\end{pmatrix}.
\label{eq:def of kaon isospinor}
\eea
As discussed in Ref.~\cite{ezoe-hosaka}, the ansatz Eq.~(\ref{ans: our ansatz}) describes the kaon fluctuation around the rotating hedgehog soliton, 
and differs from the one of Callan and Klebanov for the kaon around the static hedgehog soliton~\cite{Callan-Klebanov 1, Callan-Klebanov 2}.

Now we derive the equation of motion for the kaon field.
To do that, we first substitute our ansatz Eq.~(\ref{ans: our ansatz}) for the Lagrangian Eq.~(\ref{eq: lagrangian}), and then we expand $U_K$ up to second order of the kaon field, $K$.
As a result, we obtain the following Lagrangian for the kaon-nucleon system,  
\bea
	L = L_{SU(2)} + L_{KN},
\label{eq: obtaining lagrangians}
\eea
\bea
	L_{KN}
		&=& \left( D_{\mu}K\right)^{\dag}D^{\mu}K 
				- K^{\dag} a_{\mu}^{\dag} a^{\mu} K 
				- m_K^2 K^{\dag} K \nonum \\
		&&+ \displaystyle{\frac{1}{(e F_{\pi})^2}} \left\{ -  K^{\dag}K\mathrm{tr}\left[ \del_{\mu} \tilde{U} \tilde{U}^{\dag}, \del_{\nu} \tilde{U} \tilde{U}^{\dag}\right]^2 
			       - 2\left( D_{\mu}K\right)^{\dag}D_{\nu}K \mathrm{tr}\left( a^{\mu}a^{\nu}\right) \right.\nonumber \\
			  && \hspace{2cm} \left. - \displaystyle{\frac{1}{2}}\left( D_{\mu}K\right)^{\dag}D^{\mu}K \mathrm{tr}\left( \del_{\nu} \tilde{U}^{\dag} \del^{\nu} \tilde{U} \right) 
			       + 6\left( D_{\nu}K\right)^{\dag} \left[ a^{\nu},a^{\mu}\right]D_{\mu}K \right\} \nonum \\
		 && +  \displaystyle{\frac{3 i}{F_{\pi}^2}} B^{\mu} \left[ \left( D_{\mu} K \right)^{\dag} K - K^{\dag}  \left( D_{\mu} K \right) \right],
\label{eq:KN lagrangian}
\eea
where the covariant derivative is defined as $D_{\mu} = \del_{\mu} + v_{\mu}$, and the vector and axial vector currents are
\bea
	v_{\mu} &=& \displaystyle{\frac{1}{2}} \left( \tilde{\xi}^{\dag} \del_{\mu} \tilde{\xi} + \tilde{\xi} \del_{\mu} \tilde{\xi}^{\dag} \right), \\
	a_{\mu} &=& \displaystyle{\frac{1}{2}} \left( \tilde{\xi}^{\dag} \del_{\mu} \tilde{\xi} - \tilde{\xi} \del_{\mu} \tilde{\xi}^{\dag} \right).
\eea
In these equations, the tilded quantities are rotating;
\bea
	\tilde{U} = A \left( t \right) \xi^2 A^{\dag} \left( t \right), \ \ \ \ \tilde{\xi} = A \left( t \right) \xi A^{\dag} \left( t \right),
\eea
as required by our ansatz Eq.~(\ref{ans: our ansatz}).
Finally, the last term of Eq.~(\ref{eq:KN lagrangian}) is derived from the Wess-Zumino term with the baryonic current~\cite{ANW}, $B_{\mu}$.

Next, we decompose the kaon field into the two-component isospinor and spatial wave functions,
and expand the latter into partial waves by the spherical harmonics,  $Y_{lm} \left( \hat{r} \right)$,
\bea
	\begin{pmatrix}
		K^+ \\
		K^0
	\end{pmatrix}
	       &=& \psi_I K \left( t, \bm{r} \right) \nonum \\
	       &\rightarrow& \psi_I K \left( \bm{r} \right) \exp \left( - i E t \right),
\label{eq:decompose 1}
\eea 
\bea
	 K \left( \bm{r} \right) = \sum_{\alpha l m} C_{l m \alpha} Y_{lm} \left( \hat{r} \right) k_l^{\alpha} \left( r \right),
\label{eq:decompose 2}
\eea
where $\psi_I$ is the two component isospinor.

Finally, taking a variation with respect to the kaon wave function, we obtain the equation of motion for each partial wave, $k_l^{\alpha} \left( r \right)$,
\bea
	- \displaystyle{\frac{1}{r^2}} \displaystyle{\frac{d}{dr}} \left( r^2 h\left( r \right) \displaystyle{\frac{dk_l^{\alpha} \left( r \right)}{dr}}\right)
		- E^2 f \left( r \right) k_l^{\alpha} \left( r \right)
		+ \left( m_K^2 + V \left( r \right) \right) k_l^{\alpha} \left( r \right) = 0,
\label{eq: equation of motion}
\eea
where $h \left( r \right)$ and $f \left( r \right)$ are functions depending on the profile function, $F \left( r \right)$, and $E$ is the energy of the kaon including the rest mass of the kaon. 
The last term in Eq.~(\ref{eq: equation of motion}), $V \left( r \right)$, is the kaon-nucleon interaction term.
In Appendix~\ref{appendix}, we show explicit forms of each term in Eq.~(\ref{eq: equation of motion}).

\section{Scattering states}
\label{sec: scattering state}
In this section, we discuss phase sifts for the s- and p-wave kaon nucleon scattering states.
As we mentioned in the previous section, in this paper, we consider three parameter sets for $F_{\pi}$ and $e$  shown in the Table~\ref{tab: parameter sets}.
\begin{table}[htb]
	\begin{center}
		\caption{Parameter sets and binding energy~(B.E.)}
			\begin{tabular}{| c || c | c | c |}  \hline
						 	  & $F_{\pi}$ [MeV]& $e$	& B.E. [MeV]	\\ \hline
				Parameter set A & 205		     & 4.67	& 20.6		\\ \hline
				Parameter set B & 186		     & 4.82 & 32.2		\\ \hline
				Parameter set C & 129		     & 5.45 & 81.3		\\ \hline
			\end{tabular}
\label{tab: parameter sets}
	\end{center}
\end{table}
For the reason that we discuss later, these parameter sets reproduce the same moment of inertia such that the mass splitting of the nucleon and delta becomes the physical value.
\begin{itemize}
\item{Parameter set A}:
we employ the pion decay constant slightly larger than the physical one. 
This is motivated by that the kaon decay constant, $F_{K}$, is larger than the pion one~$\left( F_{K} = 227 \ \mathrm{MeV} \right)$, and that we are interested in a physical system of the pion and kaon.
Therefore we choose $F_{\pi} = 186 \times 1.1 = 205$ MeV.
The Skyrme parameter, $e$, is then fixed to reproduce the $N \Delta$ mass splitting together with the above $F_{\pi}$ value.

\item Parameter set B: this is adjusted to fit the $N \Delta$ mass splitting with $F_{\pi}$ fixed at the experimental value.

\item Parameter set C: this is proposed by Adkins, Nappi, and Witten~\cite{ANW}, which reproduces the observed masses of the nucleon and the delta.

\end{itemize}

For the parameter set A, there is one bound state for the $I = 0$ channel with the binding energy 20.6 MeV while, for the $I = 1$ channel, no bound state exists.
There is one bound state for the $I = 0$ and $I=1$ channels with the sets B and C
\footnote{
In our previous paper~\cite{ezoe-hosaka}, we reported that there was no bound state for the $I=1$ channel with the parameter set B.
After improving our numerical calculations, however, we have found a very shallow bound state with binding energy 0.2 MeV.
}. 
For $KN$ channel, the kaon and nucleon does not form a bound state due to the strong repulsion by the Wess-Zumino term for three parameter sets.

We have calculated phase shifts for the s- and p-wave kaon nucleon scattering states for all parameter sets. 
However, for realistic situations of kaon and nucleon systems, it turns out that the use of the physical pion decay constant is important.
Therefore, in the following discussions, we will present in most cases the results of using the parameter sets A and B.  

\begin{figure}[htb]
\begin{minipage}{0.5\hsize}
	\begin{center}
		\includegraphics[width=8cm]{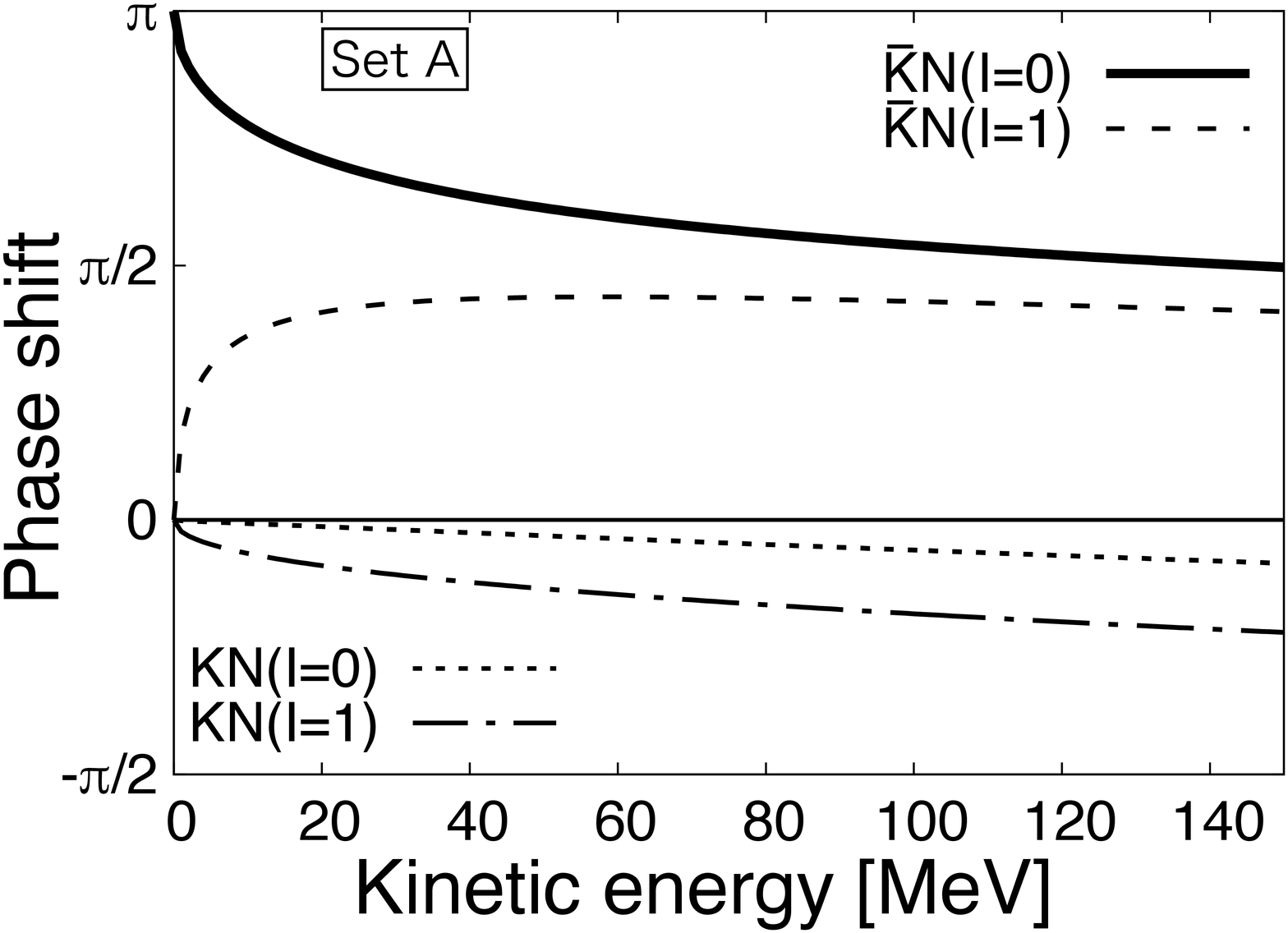}
	\end{center}
\end{minipage}%
\begin{minipage}{0.5\hsize}
 	\begin{center}
		\includegraphics[width=8cm]{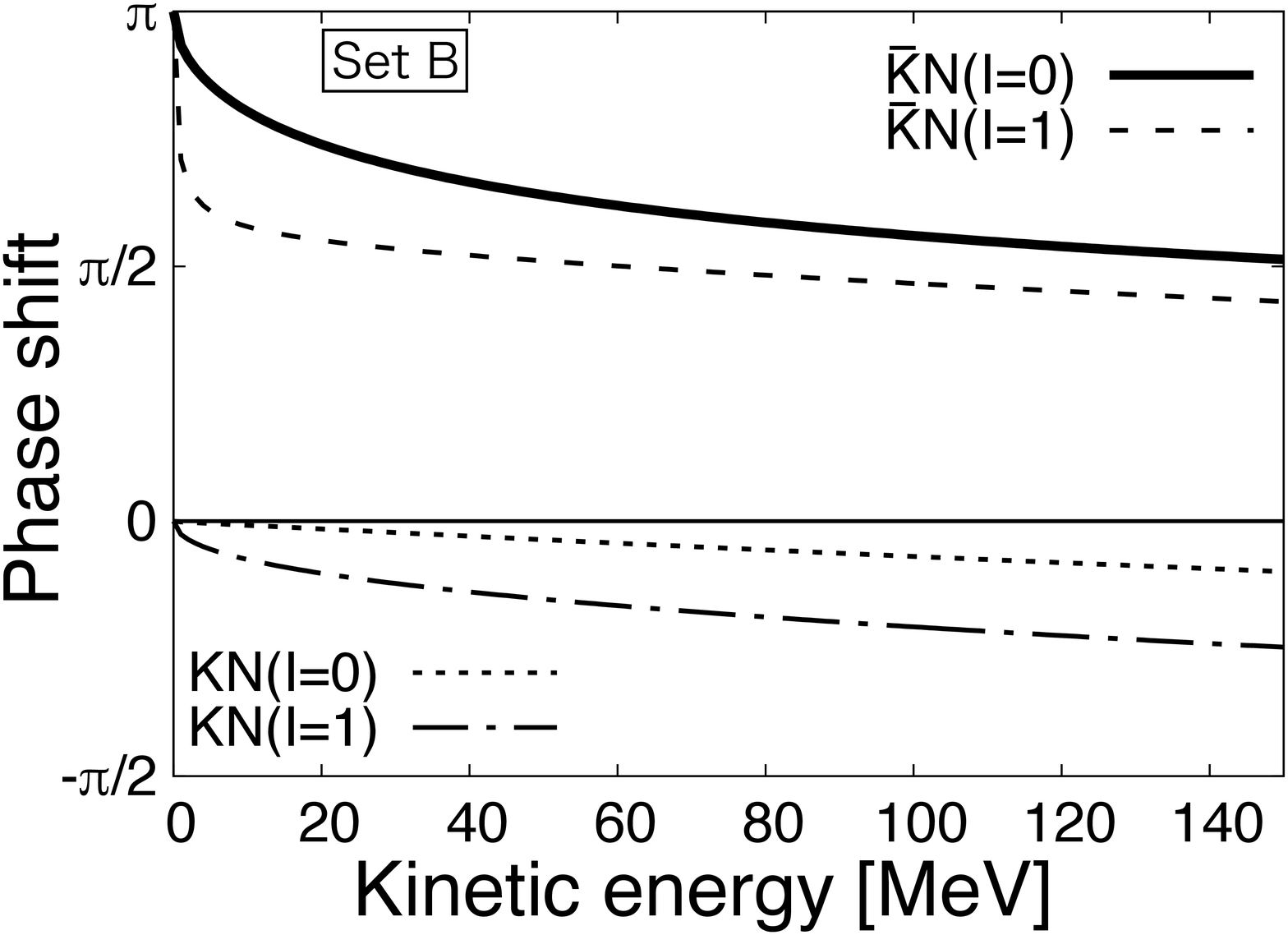}
	\end{center}
\end{minipage}
\caption{Phase shifts for the kaon-nucleon scattering state with $J^P = 1/2^-$ for the parameter sets A~(left) and B~(right).}
\label{fig: s-wave phase shift}
\end{figure}
First we show in Fig.~\ref{fig: s-wave phase shift} phase shifts for s-wave scatterings with various channels as functions of the kinetic energy $\varepsilon$, which is defined by  $E = m_K + \varepsilon$.  
For the set A~(left), the phase shift of the $\bar{K}N \left( I = 0 \right)$ channel starts from $\pi$ at $\varepsilon = 0$, reflecting the fact that there is one bound state.
For the $\bar{K}N \left( I = 1 \right)$ channel, there is not a bound state but it shows attractive nature as the positive phase shifts indicate.
For the $KN$ scattering, both $I = 0$ and $I = 1$ channels are weakly repulsive but the $I = 1$ channel is stronger. 

For the set B~(right), $\bar{K}N$ channels allow a bound state for both $I = 0$ and $I = 1$.
The bound state of $I = 1$ is, however, very shallow indicating that the attractive interaction is weaker than in the $I = 0$ channel.
The $I = 1$ bound state disappears by slightly increasing the pion decay constant as chosen in the parameter set A.  
We may further attempt fine tuning of the parameters, but we will not do this because our present model contains only $KN$ channels.
Physically, the inclusion of the $\pi \Sigma$ channels is very important, which we will do in the future.
 
From Fig.~\ref{fig: s-wave phase shift}, we find that the strength of repulsion for $K N$ and attraction for $\bar{K} N$ are stronger for the set B than for the set A.
This reflects the fact that the obtaining potential is approximately proportional to $1/F_{\pi}^2$ as the Weinberg-Tomozawa theorem implies~\cite{WT 1, WT 2}.

\begin{figure}[htb]
	\begin{center}
		\includegraphics[width=10cm]{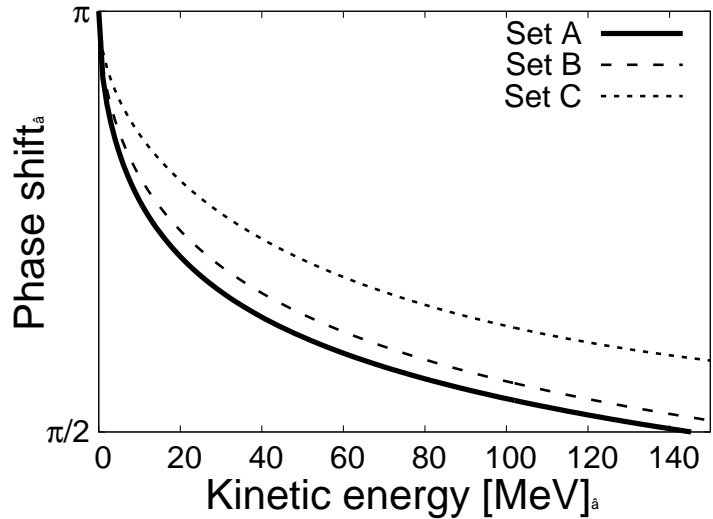}
	\end{center}
\caption{The phase shifts for the $\bar{K} N (J^P = 1/2^-, I = 0)$ channel with the three parameter sets A, B, and C.}
\label{fig: phase shift for KbarN}
\end{figure}
To complete the discussions up to here, we show the phase shifts for all parameter sets A, B, and C for the $\bar{K} N \left( J^P = 1/2^- , \ I = 0 \right)$ channel in Fig.~\ref{fig: phase shift for KbarN}.
In this figure, we can find that the attraction between the anti-kaon and nucleon becomes stronger in the order of the set A, B, and C,
which is consistent with the properties of the $\bar{K} N$ bound states shown in Table.~\ref{tab: parameter sets}.

\begin{figure}[H]
\begin{minipage}{0.5\hsize}
	\begin{center}
		\includegraphics[width=8cm]{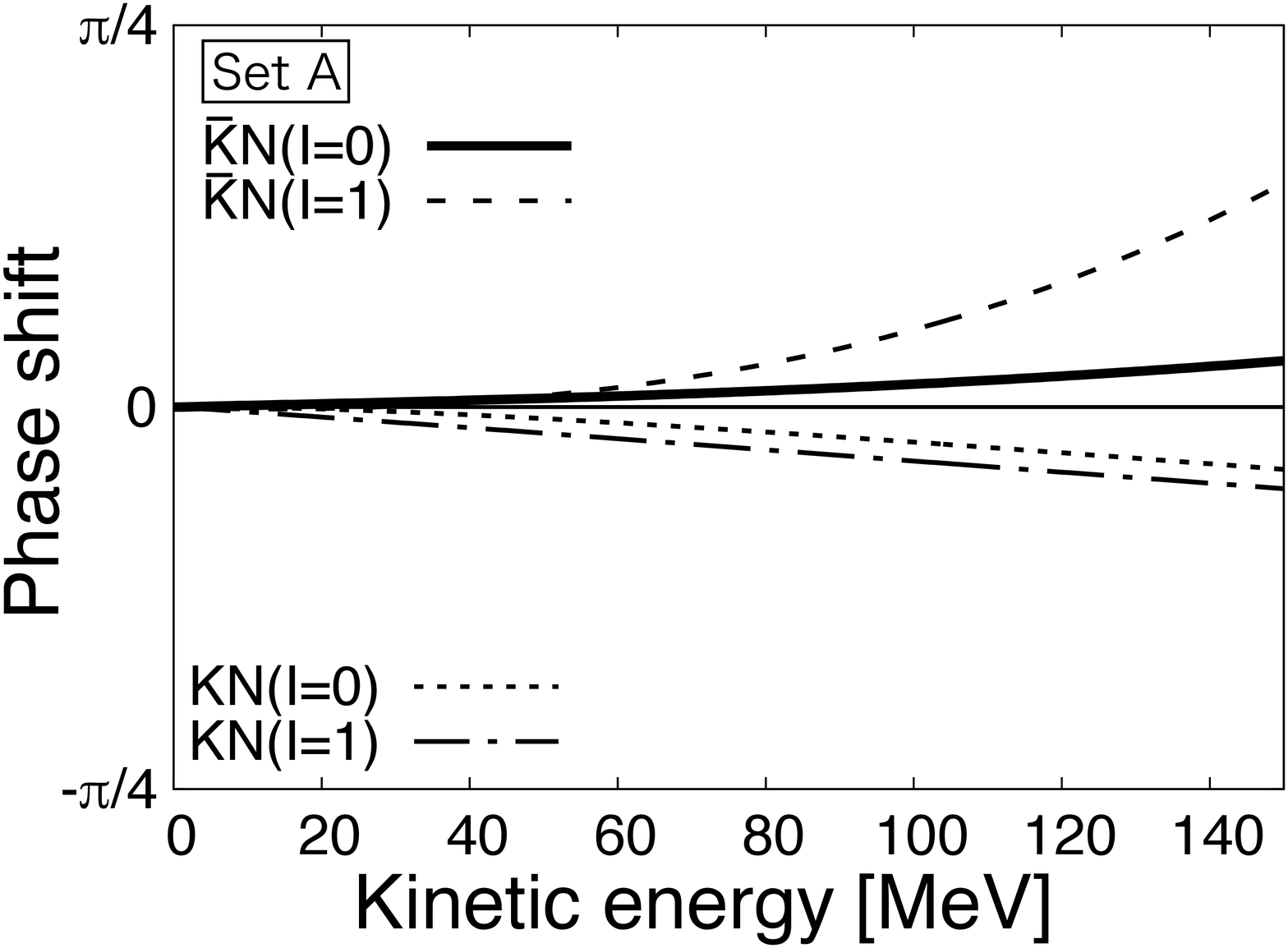}
	\end{center}
\end{minipage}%
\begin{minipage}{0.5\hsize}
 	\begin{center}
		\includegraphics[width=8cm]{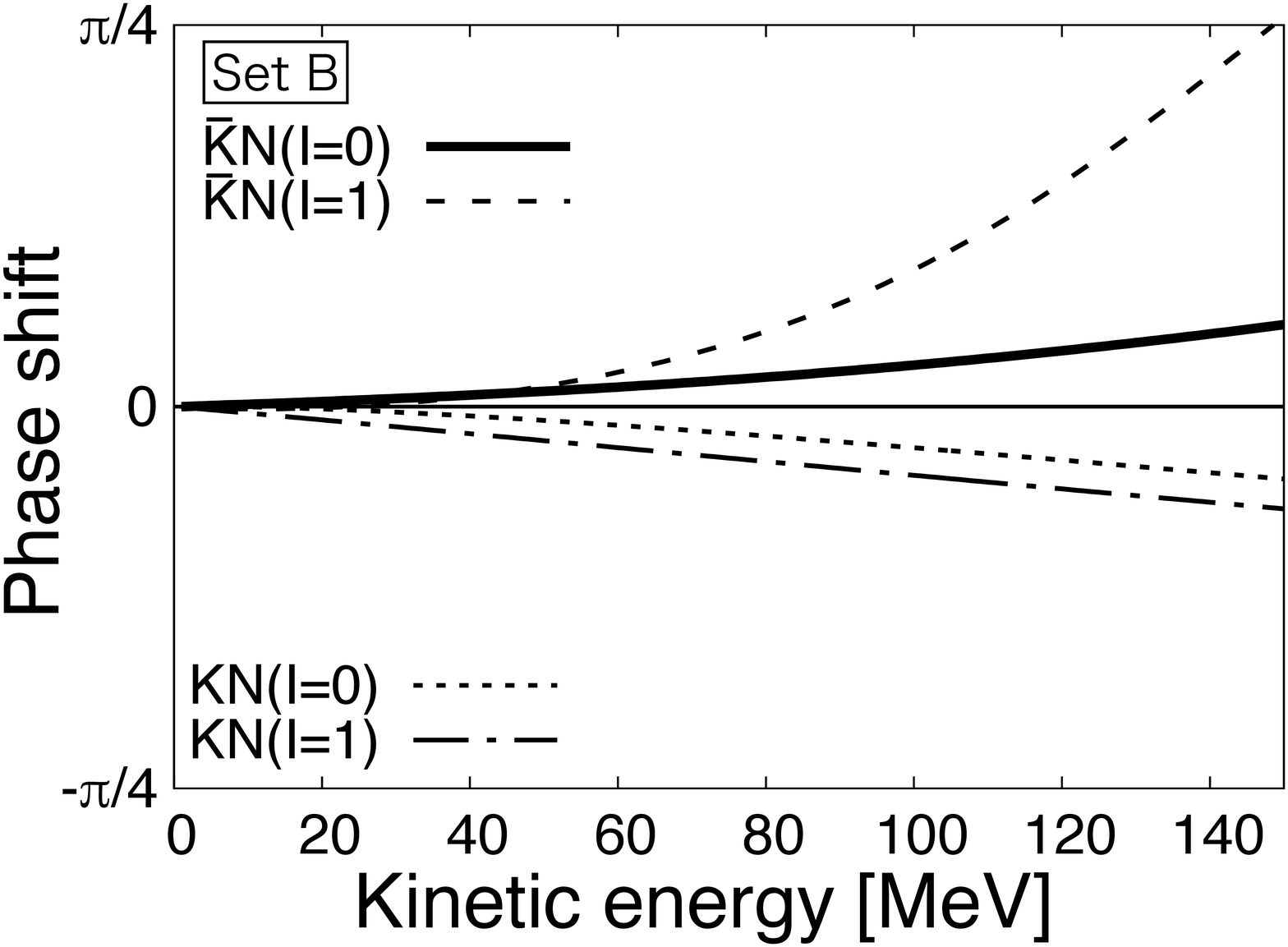}
	\end{center}
\end{minipage}
\caption{Phase shifts for the kaon-nucleon scattering state with $J^P = 3/2^+$ for the parameter sets A~(left) and B~(right).}
\label{fig: p-wave phase shift}
\end{figure}

Next, in Fig.~\ref{fig: p-wave phase shift}, we have shown the phase shifts for p-waves, first for $J^P = 3/2^+$ channels.
In both sets A and B, the phase shifts show the attractive and repulsive behaviors for $\bar{K} N$ and $KN$ channels, respectively. 
However, the strength of them are weaker than those of the s-wave.
The phase shifts in the $\bar{K} N$ channels show that the $I=1$ channel is more attractive than the $I=0$ one due to the stronger isospin-dependent LS force in the $I=1$ channel.  

For the other LS partner of $J^P = 1/2^+$ channel, the interaction shows a strong attraction as proportional to $1/r^2$.
Because of this, the system becomes unstable and physically meaningful solutions are not allowed. 
We consider that it is related to the hedgehog structure, but physical meaning is not yet clarified.  

Finally, let us evaluate the scattering length, $a$, for the $\bar{K}N \left( J^P = 1/2^- \right)$ scattering state which is defined by 
\bea
	a = - \lim_{k \to 0} \displaystyle{\frac{\tan \delta \left( k \right)}{k}},
\eea
where $k$ is the wave number and $\delta \left( k \right)$ is the phase shift.
From this equation, we have obtained $a_0 = 1.56$ fm and $a_1 = -3.38$ fm for the $\bar{K}N$ scattering with isospin 0 and 1 channels, respectively. 
As a result, we have $a_{\bar{K}N} = -0.91$ fm as the $\bar{K}N$ scattering length.
These values are larger than experimental values and other theoretical calculations~\cite{scattering length 1, scattering length 2,  scattering length 3, scattering length 4, scattering length 5}.
In the present paper, we will not make further quantitative discussions because the inclusion of the $\pi \Sigma$ channels is needed for more realistic comparison.  

\section{Potential}
\label{sec: potential}
In this section, we investigate the $\bar{K}N$ potential in detail.
Numerical results are then fitted to a simple functional form which are useful for various applications to the study of $\bar{K}$-nucleon systems.
First, we consider the potential for the parameter set A.
For the sets B and C, we discuss them with scaling rules from the set A to the sets B and C.     

\subsection{Derivation and classification of the potential}
Let us start with the equation of motion Eq.~(\ref{eq: equation of motion}) in the following Schr\"{o}dinger-like form with the potential $U \left( r \right)$ in units of MeV,
\bea
	- \displaystyle{\frac{1}{m_K + E}} \displaystyle{\frac{1}{r^2}} \frac{d}{dr} \left( r^2 \frac{dk_l^{\alpha} \left(r\right)}{dr} \right)
		+ U \left( r \right) k_l^{\alpha} \left(r\right)
		= \varepsilon k_l^{\alpha} \left(r\right),
\label{eq:schrodinger equation}
\eea
where $E = m_K + \varepsilon$, 
and
\bea
	U \left( r \right)
		&=& - \displaystyle{\frac{1}{m_K + E}} \left[ \displaystyle{\frac{h \left( r \right)-1}{r^2}} \displaystyle{\frac{d}{dr}} \left( r^2 \displaystyle{\frac{d}{dr}} \right)
			+ \displaystyle{\frac{d h(r)}{dr}} \displaystyle{\frac{d}{dr}} \right]
			- \displaystyle{\frac{\left( f \left( r \right) - 1 \right) E^2}{m_K + E}}
			+ \displaystyle{\frac{V (r)}{m_K + E}}. \nonum \\
\label{eq:effective potential}
\eea
The potential $U \left( r \right)$ has the following properties~\cite{ezoe-hosaka};
it is nonlocal and depends on the energy of the kaon.
Second, it contains isospin independent and dependent central terms and spin-orbit~(LS) terms.
Finally, there are repulsive components proportional to $1/r^2$ at short distances.

Because this expression contains the derivative operators, we define the equivalent local potential $\tilde{U} \left( r \right)$ with the kaon partial wave function, $k_l^{\alpha} \left(r\right)$,
\bea
	\tilde{U} \left( r \right) \equiv \displaystyle{\frac{U \left( r \right) k_l^{\alpha} \left(r\right)}{k_l^{\alpha} \left(r\right)}}.
\label{eq:potential value}
\eea
This definition, however, can not be used when the wave function becomes zero at nodal points.
We may avoid this problem by using a bound state for the isospin $I = 0$ $\bar{K}N$ channel which allows one bound state, 
and for other channels by using a scattering state with a small energy such that the first node of the wave function appears at a large $r$ where the potential is sufficiently suppressed.  
In the following, we show the results for the scattering energy $\varepsilon = 27$~MeV, while we have confirmed that results do not change as long as the scattering energy is small.  

As mentioned already, the equivalent local potential $\tilde{U} \left( r \right)$ has four kinds of components.
Here, we decompose them further into seven components reflecting different origins of the potential, 
\bea
	\tilde{U} \left( r \right) 
		&=& \tilde{U}_0^c \left( N, r \right) + \tilde{U}_{\tau}^c \left( N, r \right) I_{KN}  \nonum \\
			&& + \tilde{U}_0^{LS} \left( N, r \right) J_{KN} + \tilde{U}_{\tau}^{LS} \left( N, r \right) J_{KN} I_{KN} \nonum \\
			&& + \tilde{U}_0^c \left( WZ, r \right) + \tilde{U}_0^{LS} \left( WZ, r \right) J_{KN} \nonum \\
			&& + \tilde{U}_{l} \left( r \right),
\label{eq: more general form of our potential}
\eea
where superscripts, $c$ and $LS$, stand for the central and spin-orbit~(LS) forces, respectively, 
and 
subscripts, $0$ and $\tau$ are for isospin independent and dependent components, respectively.
The arguments, $N$ and $WZ$, indicate the terms derived from the normal terms of the Skyrme Lagrangian and the Wess-Zumino term, respectively.
In Eq.~(\ref{eq: more general form of our potential}), we have defined $I_{KN}$ and $J_{KN}$ as $I_{KN} = \bm{I}^K \cdot \bm{I}^N$ and $J_{KN} = \bm{L}^K \cdot \bm{J}^N$, respectively.
The former, $I_{KN}$, corresponds to the product of the isospin operator for the kaon and the nucleon 
and the latter, $J_{KN}$, to the product of the angular momentum of the kaon and the spin of the nucleon.
The last term in Eq.~(\ref{eq: more general form of our potential}), $\tilde{U}_{l} \left( r \right)$, is the centrifugal force of the kaon.
Because the Wess-Zumino term corresponds physically to the $\omega$-meson exchange, that is the isoscalar particle exchange~\cite{WZ term 1, WZ term 2, WZ term 3}, 
it has no isospin dependent contributions in Eq~(\ref{eq: more general form of our potential}).

The seven potential components have energy dependence, for which we make a linear approximation in terms of $\Delta_{E} \equiv \varepsilon/2m_K$, 
\bea
	\tilde{U} \left( r \right)
		&\simeq& \tilde{U} \left( r \right) + \displaystyle{\frac{\del \tilde{U} \left( r \right)}{\del \varepsilon}} \Delta_{E} \nonum \\
		&\equiv& u \left( r \right) + v \left( r \right) \Delta_{E}.
\label{eq: expanded potential}
\eea
We then fit all the components of $u \left( r \right)$ and $v \left( r \right)$ by several Gaussian type functions, 
\bea 
\label{eq: fitting function 1} 
	G_{-2} \left( r \right) &=& C_{-2} \displaystyle{\frac{1}{r^2/{R_{-2}}^2}} \exp \left( - \displaystyle{\frac{r^2}{{R_{-2}}^2}} \right) \\
\label{eq: fitting function 2} 
	G_{0} \left( r \right) &=& C_{0} \exp \left( - \displaystyle{\frac{r^2}{{R_{0}}^2}} \right)\\
\label{eq: fitting function 3} 
	G_{2} \left( r \right) &=& C_{2} \displaystyle{\frac{r^2}{{R_{2}}^2}} \exp \left( - \displaystyle{\frac{r^2}{{R_{2}}^2}} \right),
\eea
as summarized in Table~\ref{tab: fitting functionals}.
\begin{table}[htb]
	\begin{center}
		\begin{tabular}{| c || c | c |} \hline
		    				& Isospin 	& Fitting function                                                              	         				\\ \hline \hline
			Central	 	& indep.   	& $u_0^c \left( N, r \right) + v_0^c \left( N, r \right) \Delta_{E}$           	     			\\              
                    		    		&              	& $G_{-2} \left( r \right) + G_{0} \left( r \right) + G_{2} \left( r \right)$ 				\\ 
	                    		    	&		& $u_0^c \left( WZ, r \right) + v_0^c \left( WZ, r \right) \Delta_{E}$				\\
			   		    	&		& $G_{0}^{(1)} \left( r \right) + G_{0}^{(2)} \left( r \right)$						\\ \cline{2-3}
						& dep.      	& $u_{\tau}^c \left( N, r \right) + v_{\tau}^c \left( N, r \right) \Delta_{E}$ 			\\ 
               				    	&              	& $G_{0} \left( r \right) + G_{2} \left( r \right)$      		                      	 	 	\\ \hline
                    	LS    		    	& indep.   	& $u_0^{LS} \left( N, r \right)+ v_0^{LS} \left( N, r \right) \Delta_{E}$			 	\\
                    		   		&              	& $G_{0}^{(1)} \left( r \right) + G_{0}^{(2)} \left( r \right)$ 				   		\\
 	                   		  	&		& $u_0^{LS} \left( WZ, r \right) + v_0^{LS} \left( WZ, r \right) \Delta_{E}$			\\
						&		& $G_{0}^{(1)} \left( r \right) + G_{0}^{(2)} \left( r \right)$						\\ \cline{2-3}
						& dep.      	& $u_{\tau}^{LS} \left( N, r \right) + v_{\tau}^{LS} \left( N, r \right) \Delta_{E}$		\\
               			  		&              	& $G_{-2}^{(1)} \left( r \right) + G_{-2}^{(2)} \left( r \right)$ 				  	 	\\ \hline
			Centrifugal
			force			&		& $u_l \left( r \right) + v_l \left( r \right) \Delta_{E}$							\\
						&		& $ \left[ G_{0}^{(1)} \left( r \right) + G_{0}^{(2)} \left( r \right) + 1 \right] /2m_K r^2$	\\ \hline
		\end{tabular}
	\end{center}
\caption{
Various components of the $\bar{K}N$ potential and the corresponding fitted functions.
The functions $u \left( r \right)$ and $v \left( r \right)$ are for energy independent and dependent components with upper and lower indices as explained in the text.
The fitted functions $G_n \left( r \right)$ are also defined in the text, Eq.~(\ref{eq: fitting function 1}), (\ref{eq: fitting function 2}), and~(\ref{eq: fitting function 3}).
Superscripts, $(1)$ and $(2)$ of $G_{n}^{(\cdot)} \left( r \right)$, indicate that fitting parameters are different for the functions.
}
\label{tab: fitting functionals}
\end{table}
For example, the isospin independent components of the central terms derived from the normal Skyrme Lagrangian, $u_0^c \left( N, r \right)$ and $v_0^c \left( N, r \right)$, 
are fitted by the three functions,
$G_{-2} \left( r \right)$, $G_{0} \left( r \right)$, and $G_{2} \left( r \right)$.  
The first one, $G_{-2} \left( r \right)$, is the Gaussian divided by $r^2$ which is needed to reproduce a repulsive behavior in the short range, 
and the second and third are the Gaussians with polynomial of $r^0$ and $r^2$.

For the centrifugal term, we have fitted as follows,
\bea
	\tilde{U}_{l} \left( r \right) = \displaystyle{\frac{l \left( l + 1 \right)}{2 m_K r^2}} \left[ G_{0}^{(1)} \left( r \right) + G_{0}^{(2)} \left( r \right) + 1 \right],
\label{eq: centrifugal term}
\eea
where $l$ and $m_K$ are the angular momentum and the mass of the kaon, respectively.  
At short and middle distances, the centrifugal term deviates from the ordinary one of $1/r^2$ due to background fields of the hedgehog soliton.
However, at long distances, it reduces to the ordinary form.

\subsection{Numerical fitting}
\label{sec: numerical fitting}
In this subsection, we compare numerically obtained potentials with those fitted by the Gaussian forms for each component in Fig.~\ref{fig: comparison u_0^c}~--~\ref{fig: comparison centrifugal term}. 
The fitting parameters are shown in Appendix~\ref{app: fitting parameters}.

We have treated both ranges, $R_i$, and strengths, $C_i$, as fitting parameters.
Practically, we have performed the fitting as follows; 
first, we have fitted both range and strength parameters for the energy independent components because they are dominant contributions.
Then, we have determined the strength parameters of the energy dependent components with the same range parameters as the energy independent ones.      
This is because we consider that the energy independent and dependent components have the same physical origin, if based on a boson exchange picture.  

Let us now make detailed discussions for each component below. 
We concentrate on the $\bar{K} N$ potentials but we can estimate the $KN$ ones with taking into account the difference of the quantum numbers.
However, due to the nonlocality, we need to solve the equation of motion to derive more accurate potentials. 

\begin{figure}[H]
\begin{minipage}{0.5\hsize}
	\begin{center}
		\includegraphics[width=8cm]{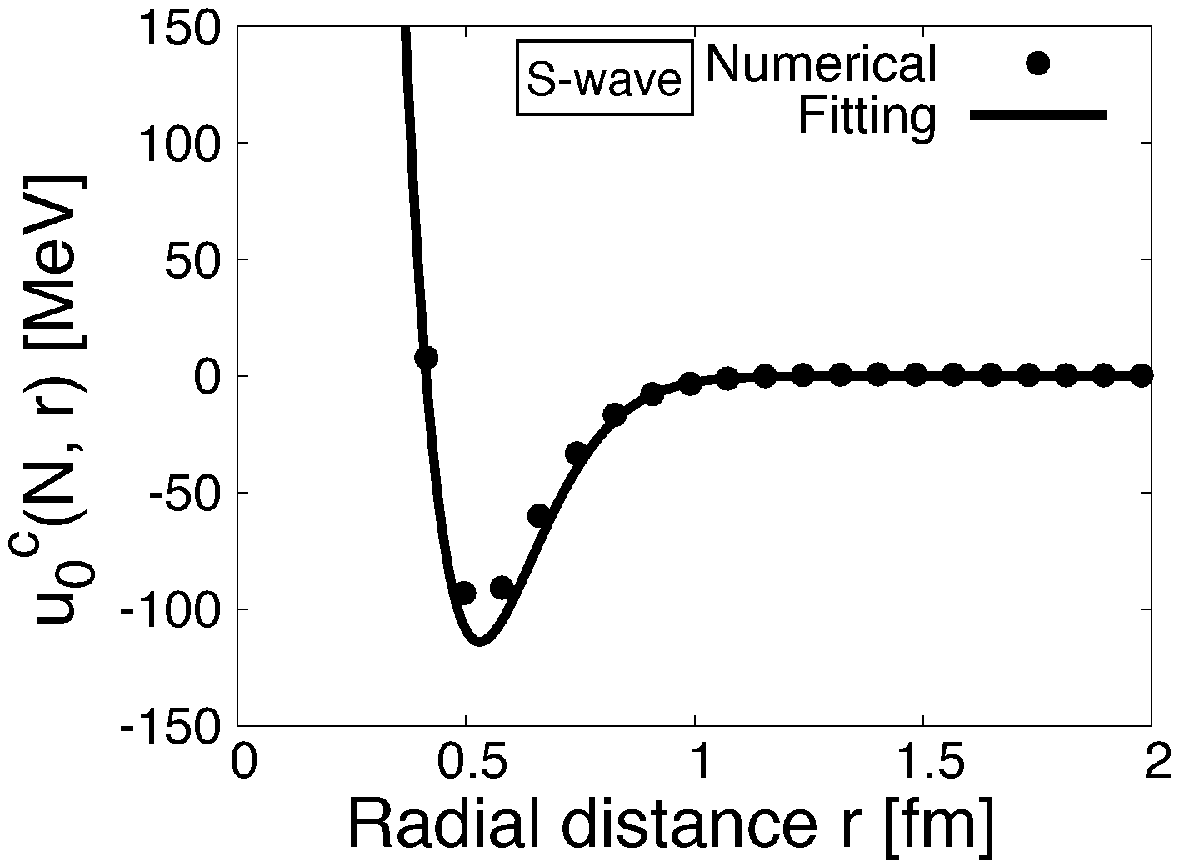}
	\end{center}
\end{minipage}%
\begin{minipage}{0.5\hsize}
 	\begin{center}
		\includegraphics[width=8cm]{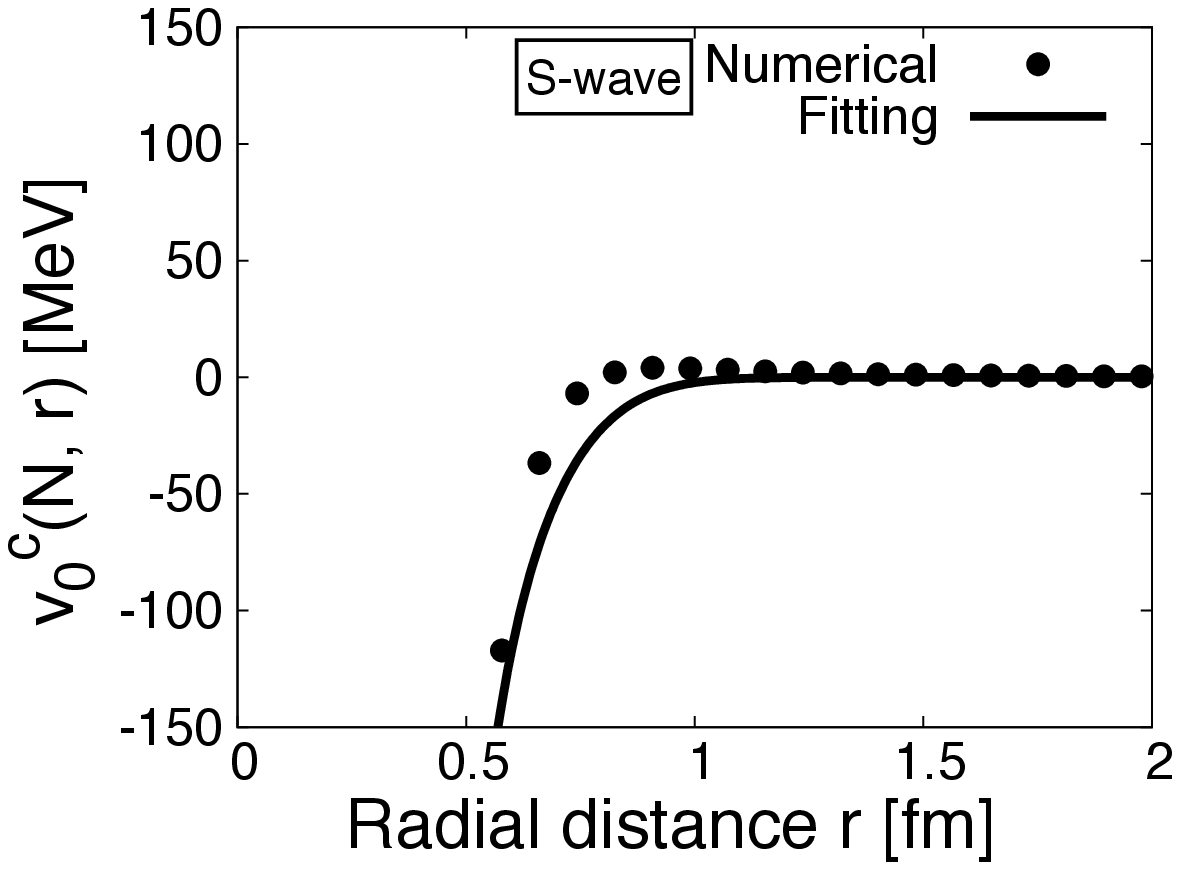}
	\end{center}
\end{minipage}
\begin{minipage}{0.5\hsize}
	\begin{center}
		\includegraphics[width=8cm]{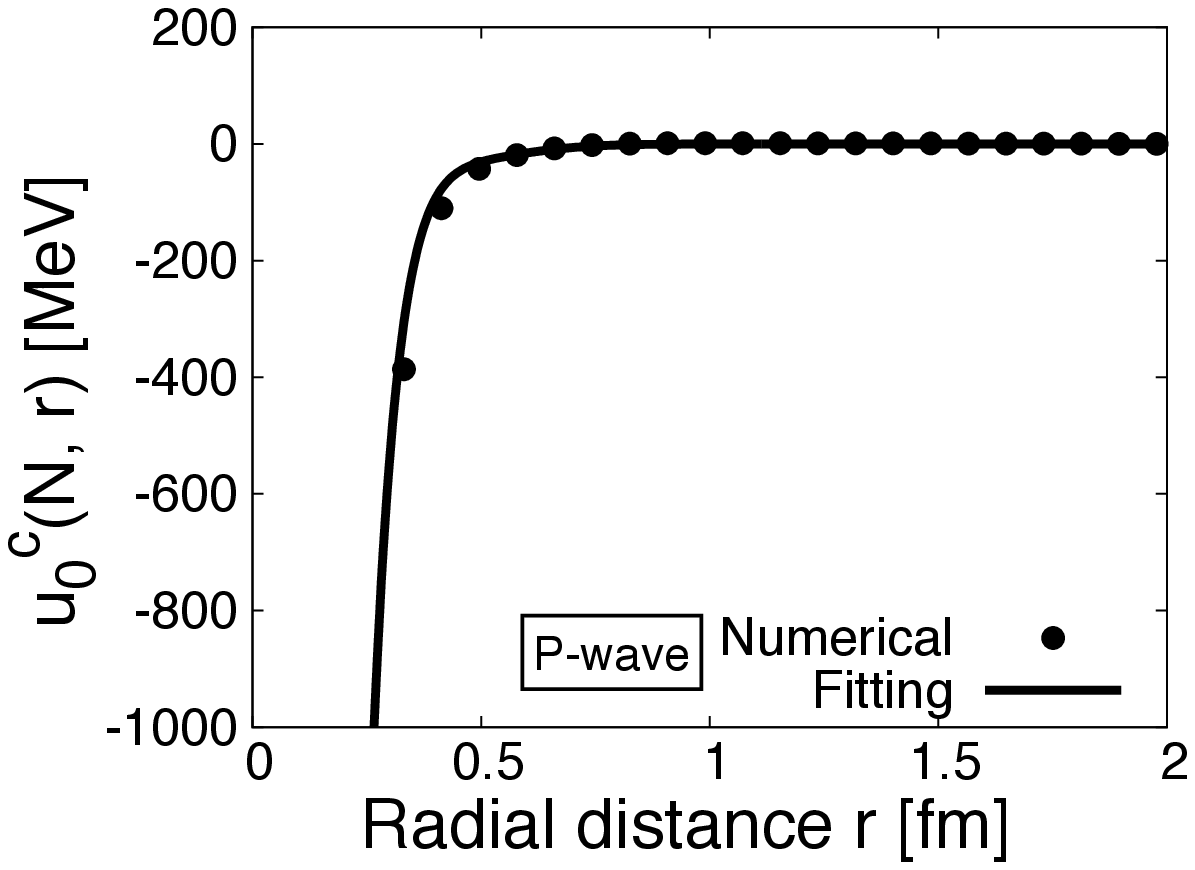}
	\end{center}
\end{minipage}%
\begin{minipage}{0.5\hsize}
 	\begin{center}
		\includegraphics[width=8cm]{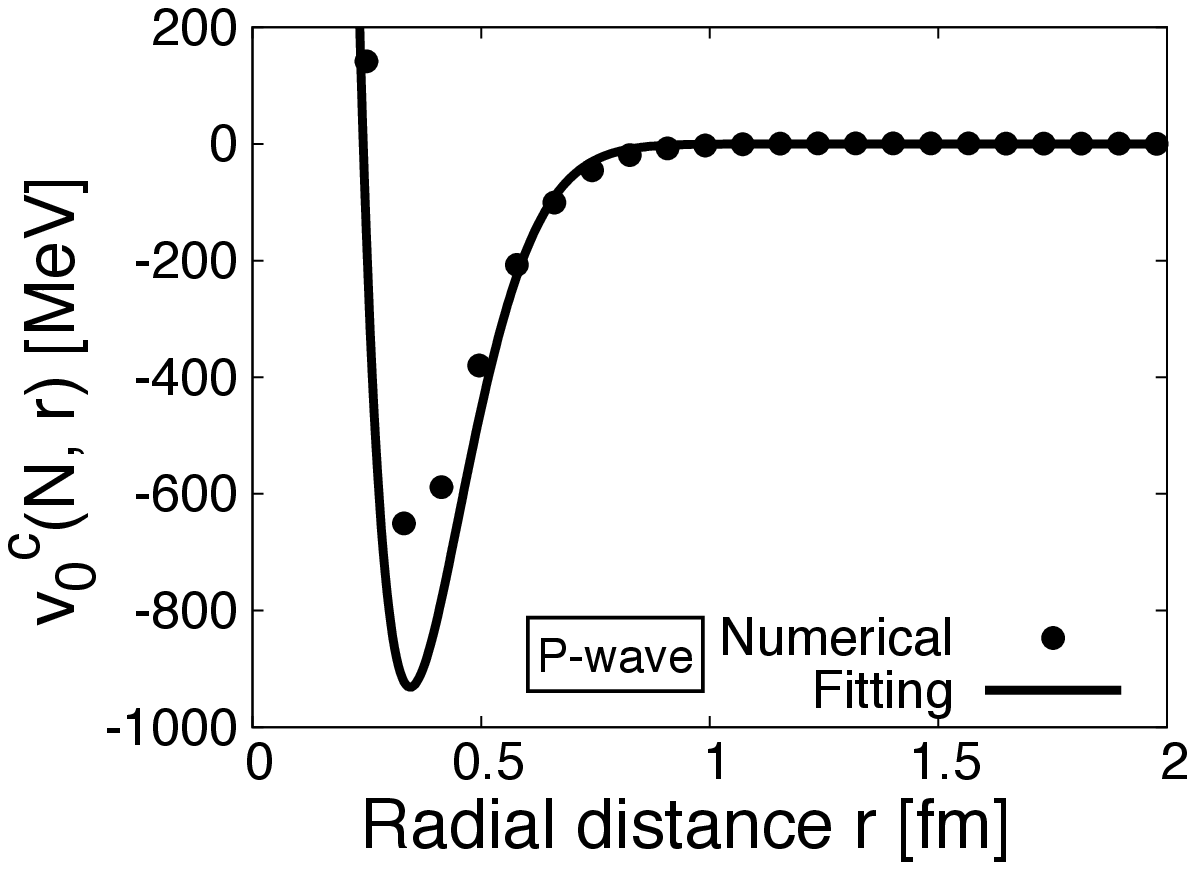}
	\end{center}
\end{minipage}
\caption{Comparisons between the numerically obtained and the Gaussian-fitted potentials 
	   for $u_0^c \left(N, r \right)$~(left) and $v_0^c \left(N, r \right)$~(right)
	   for the s-wave~(upper panels) and p-wave~(lower panels) components.}
\label{fig: comparison u_0^c}
\end{figure}
\begin{figure}[H]
\begin{minipage}{0.5\hsize}
	\begin{center}
		\includegraphics[width=8cm]{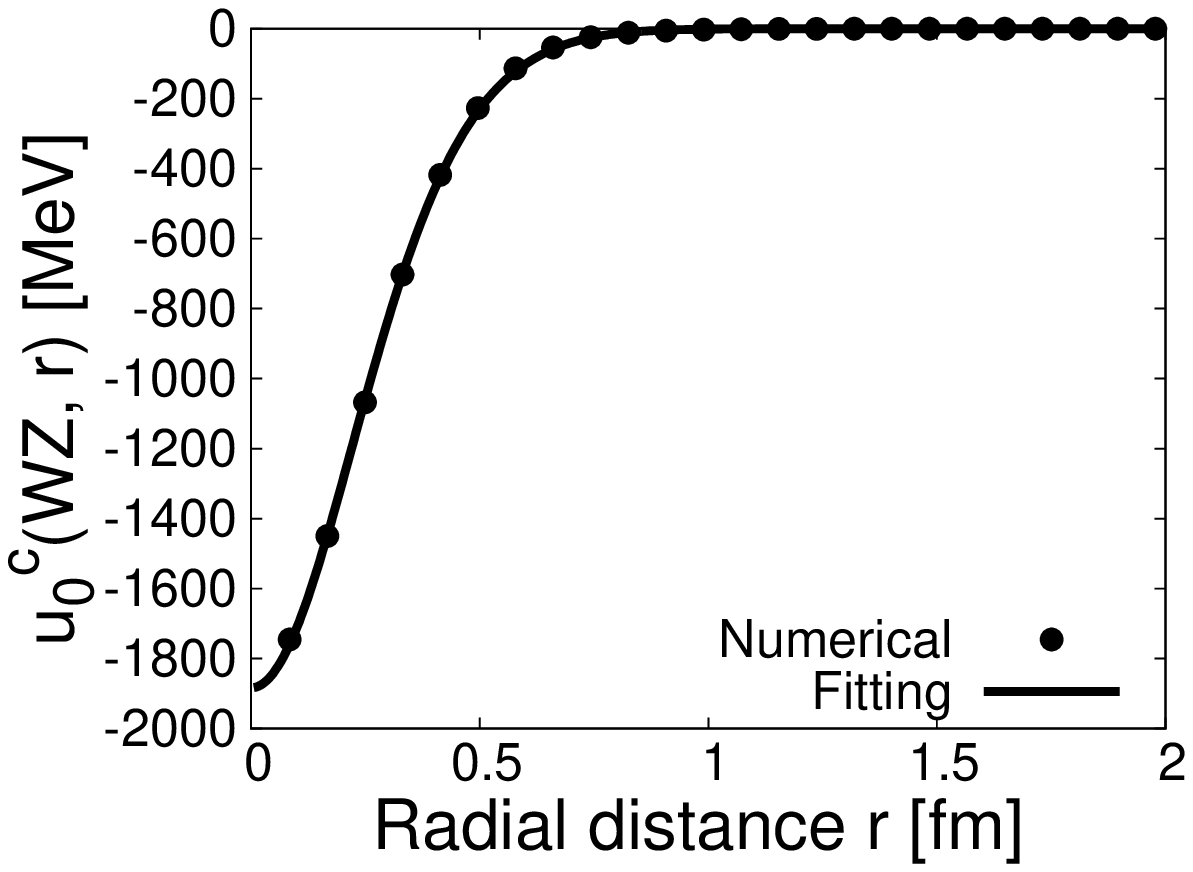}
	\end{center}
\end{minipage}%
\begin{minipage}{0.5\hsize}
 	\begin{center}
		\includegraphics[width=8cm]{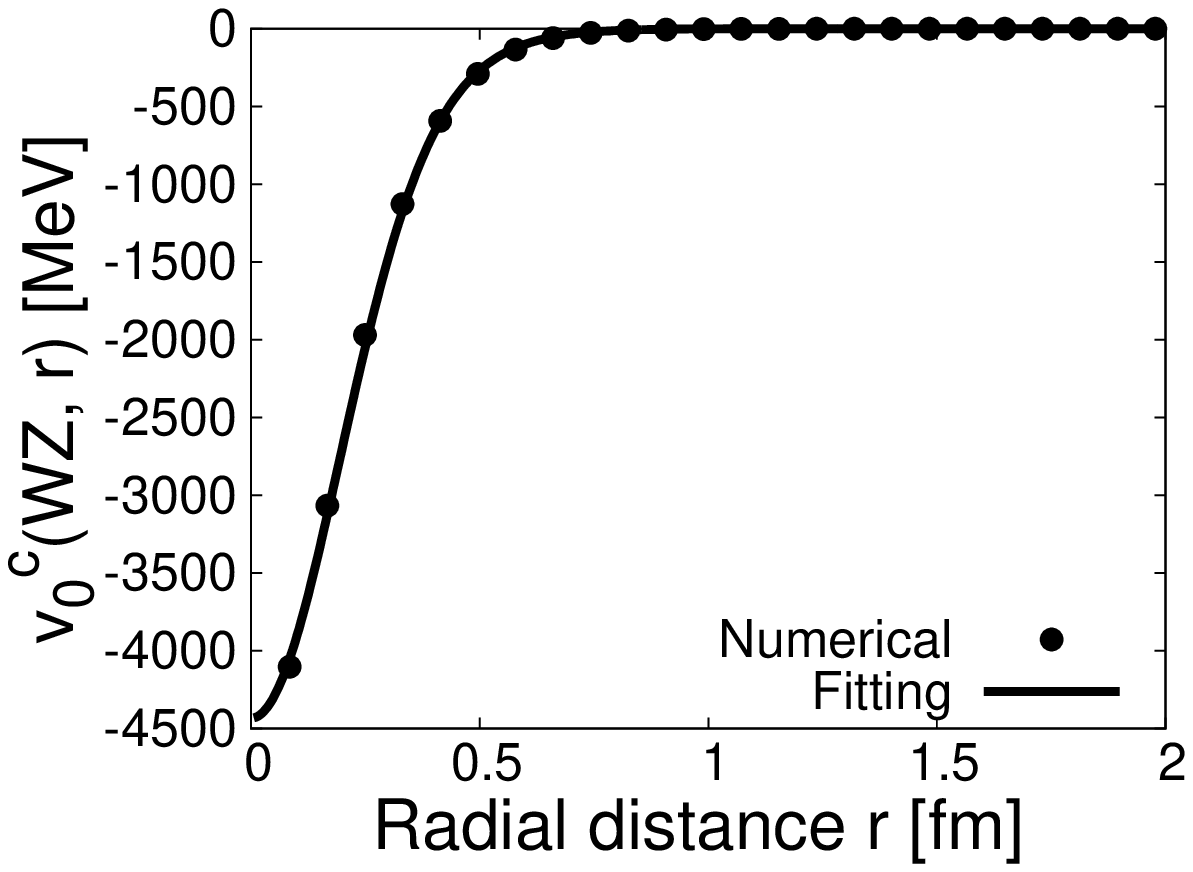}
	\end{center}
\end{minipage}
\caption{Comparisons between the numerically obtained and the Gaussian-fitted potentials 
	   for $u_0^c \left(WZ, r \right)$~(left) and $v_0^c \left(WZ, r \right)$~(right).}
\label{fig: comparison u_0^c(WZ)}
\end{figure}
\begin{figure}[H]
\begin{minipage}{0.5\hsize}
	\begin{center}
		\includegraphics[width=8cm]{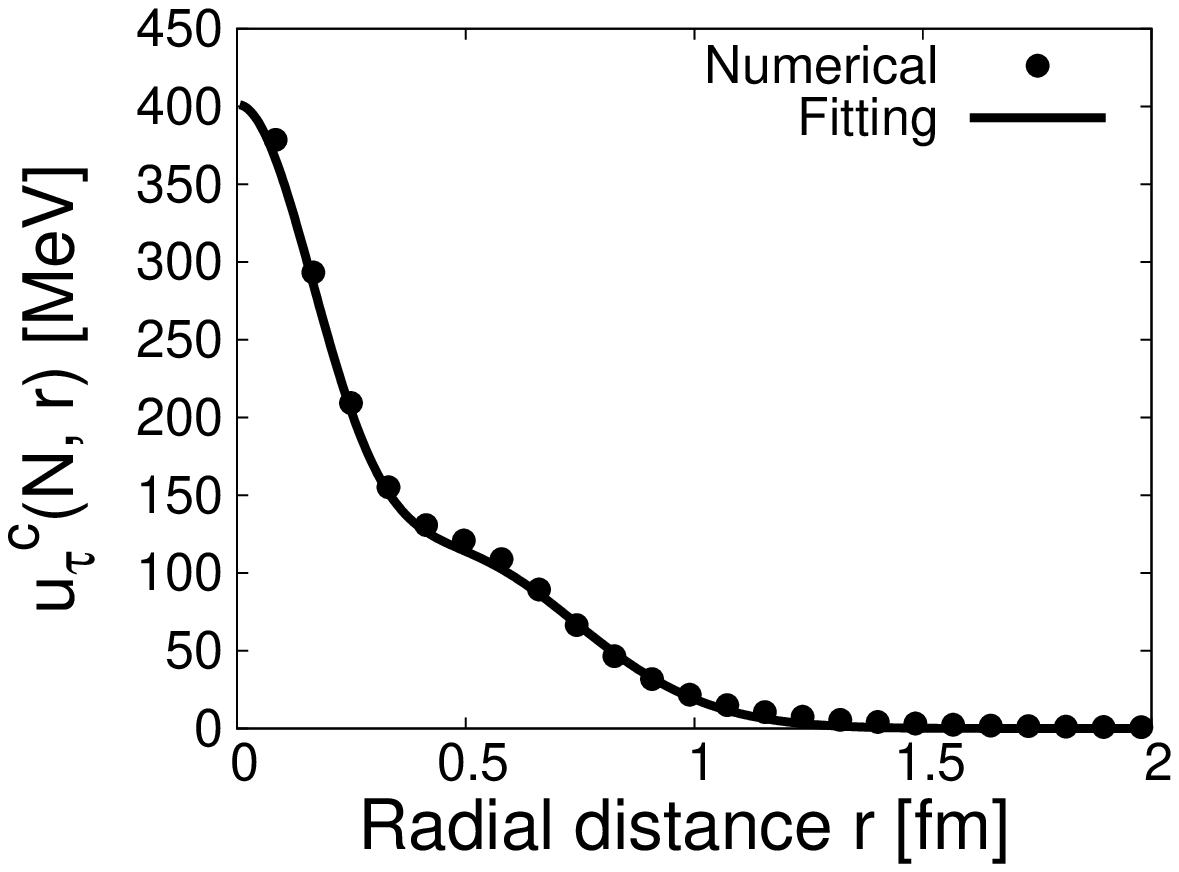}
	\end{center}
\end{minipage}%
\begin{minipage}{0.5\hsize}
 	\begin{center}
		\includegraphics[width=8cm]{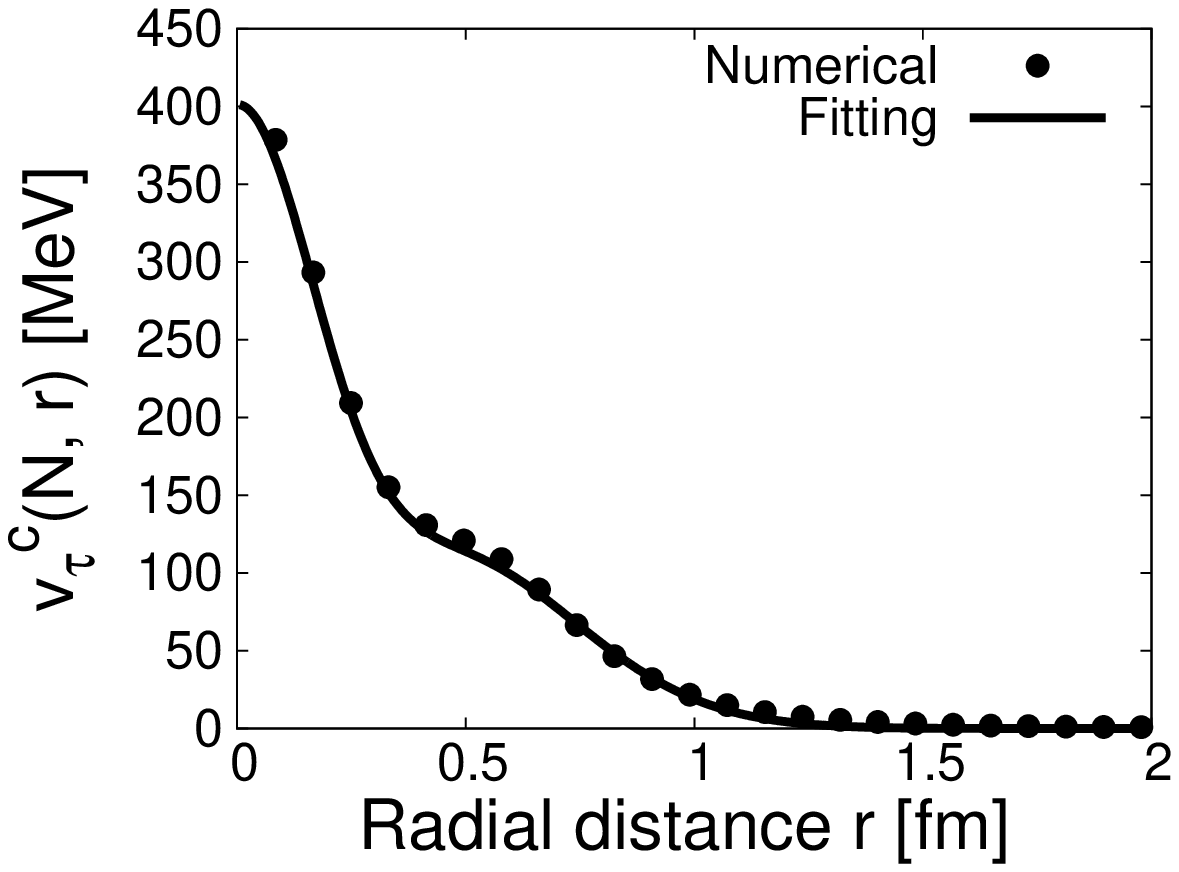}
	\end{center}
\end{minipage}
\caption{Comparisons between the numerically obtained and the Gaussian-fitted potentials 
	   for $u_{\tau}^c \left(N, r \right)$~(left) and $v_{\tau}^c \left(N, r \right)$~(right).}
\label{fig: comparison u_tau^c}
\end{figure}
\begin{figure}[H]
\begin{minipage}{0.5\hsize}
	\begin{center}
		\includegraphics[width=8cm]{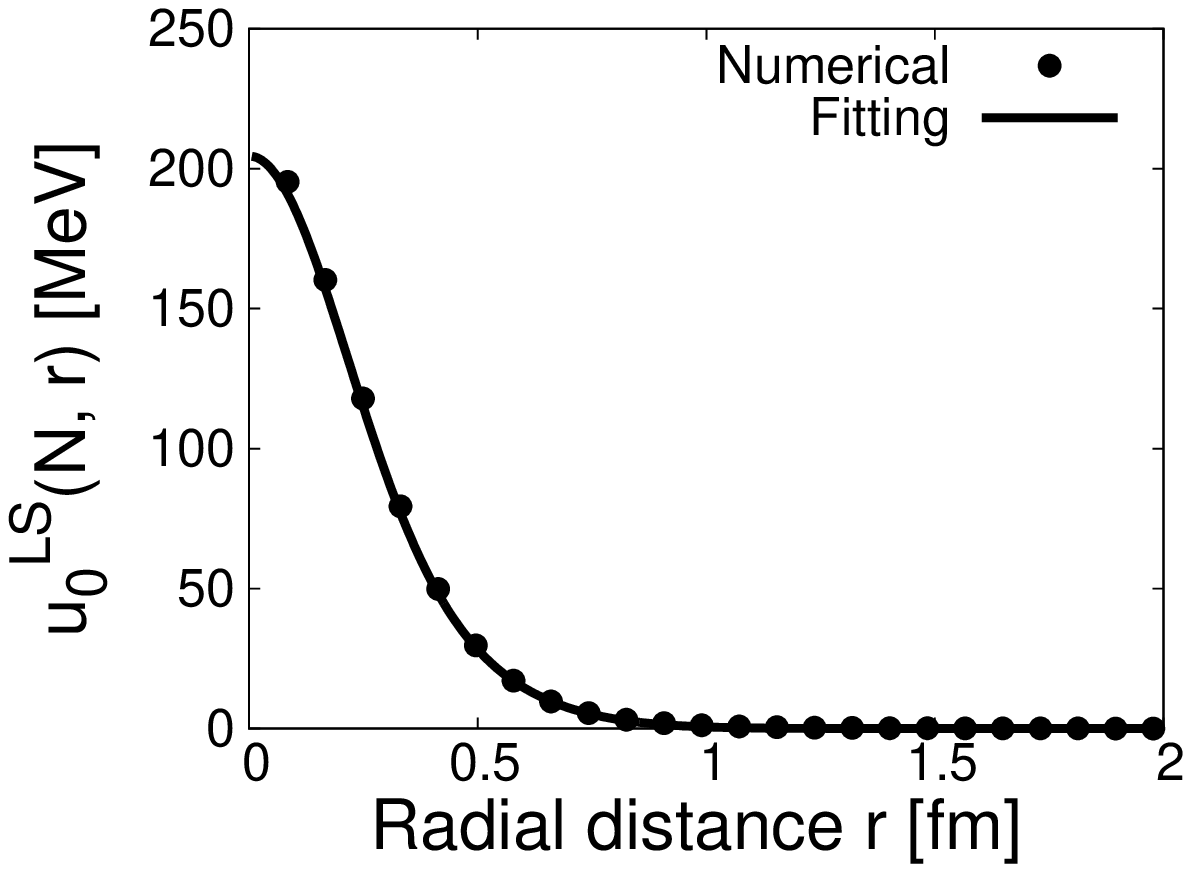}
	\end{center}
\end{minipage}%
\begin{minipage}{0.5\hsize}
 	\begin{center}
		\includegraphics[width=8cm]{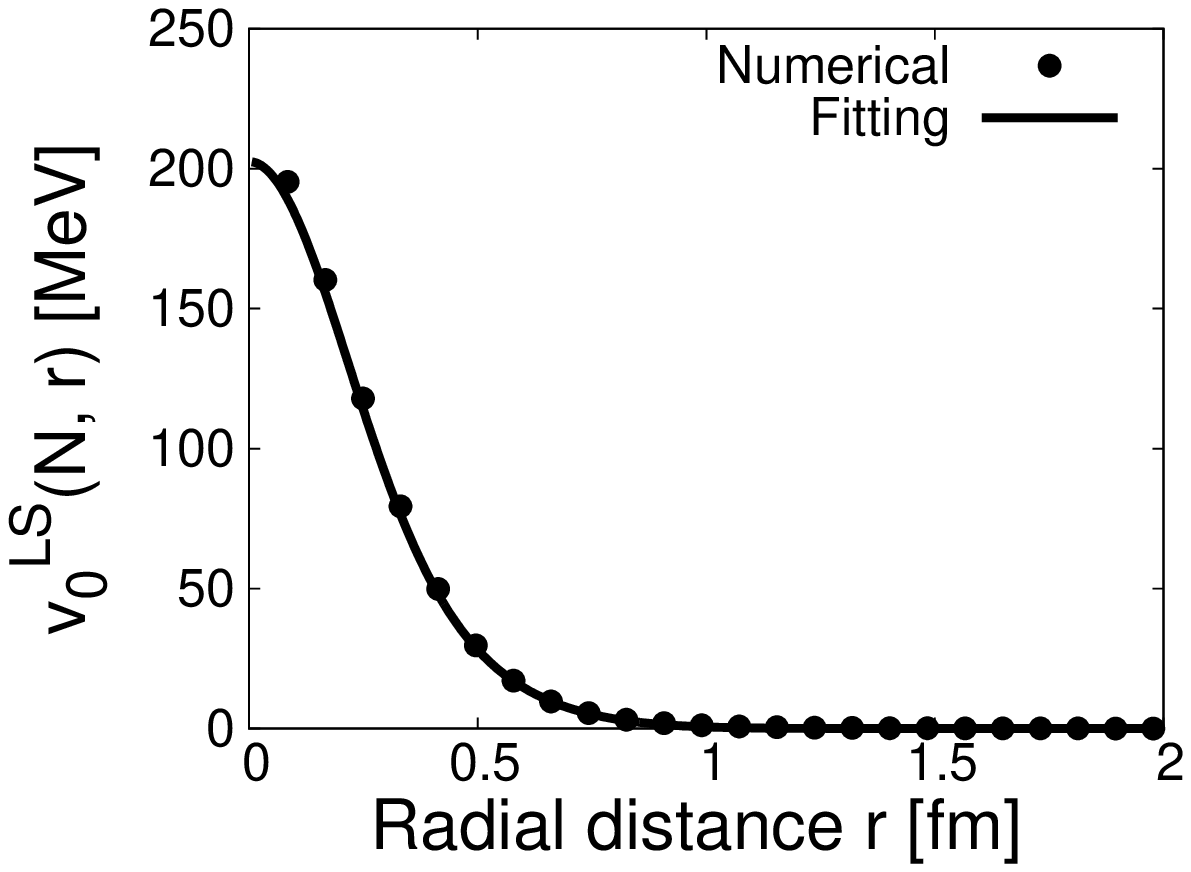}
	\end{center}
\end{minipage}
\caption{Comparisons between the numerically obtained and the Gaussian-fitted potentials 
	   for $u_0^{LS} \left(N, r \right)$~(left) and $v_0^{LS} \left(N, r \right)$~(right).}
\label{fig: comparison u_0^LS}
\end{figure}
\begin{figure}[H]
\begin{minipage}{0.5\hsize}
	\begin{center}
		\includegraphics[width=8cm]{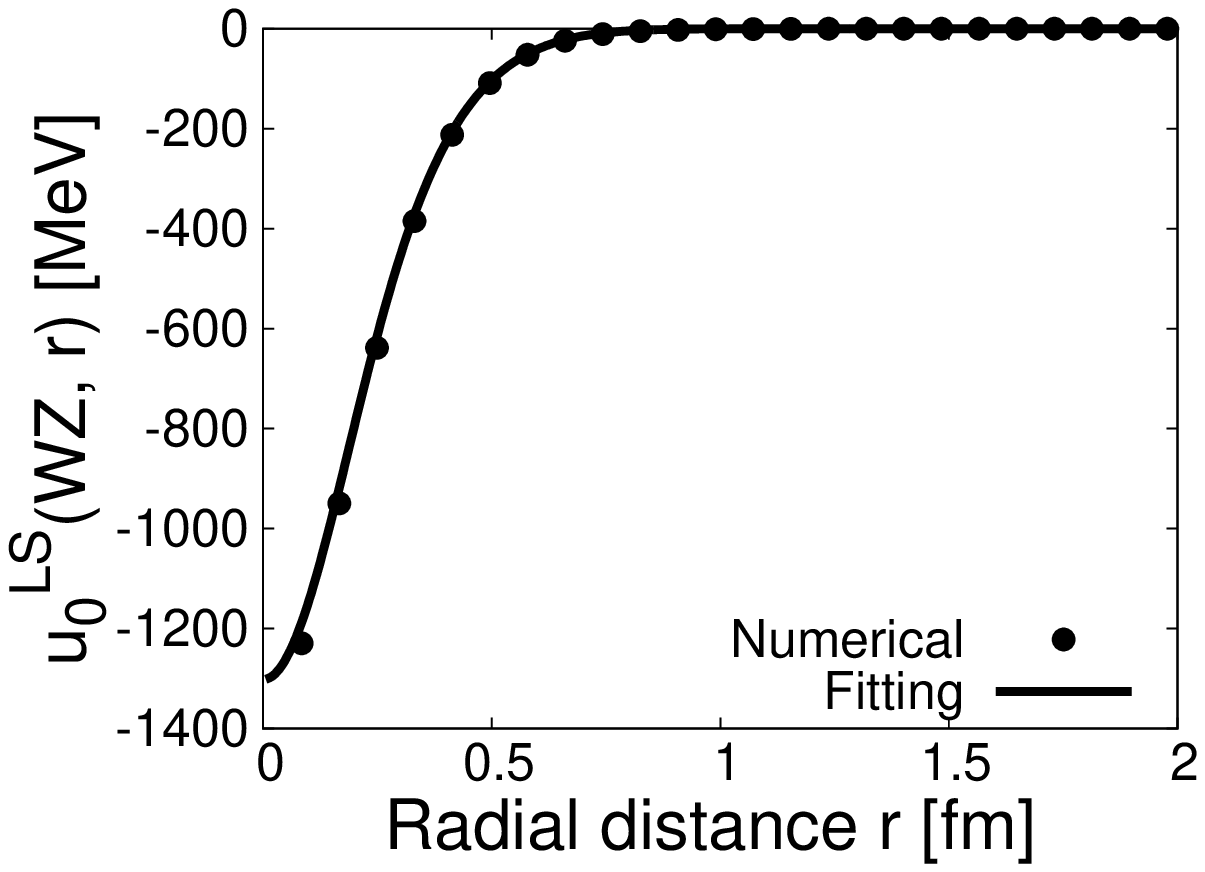}
	\end{center}
\end{minipage}%
\begin{minipage}{0.5\hsize}
 	\begin{center}
		\includegraphics[width=8cm]{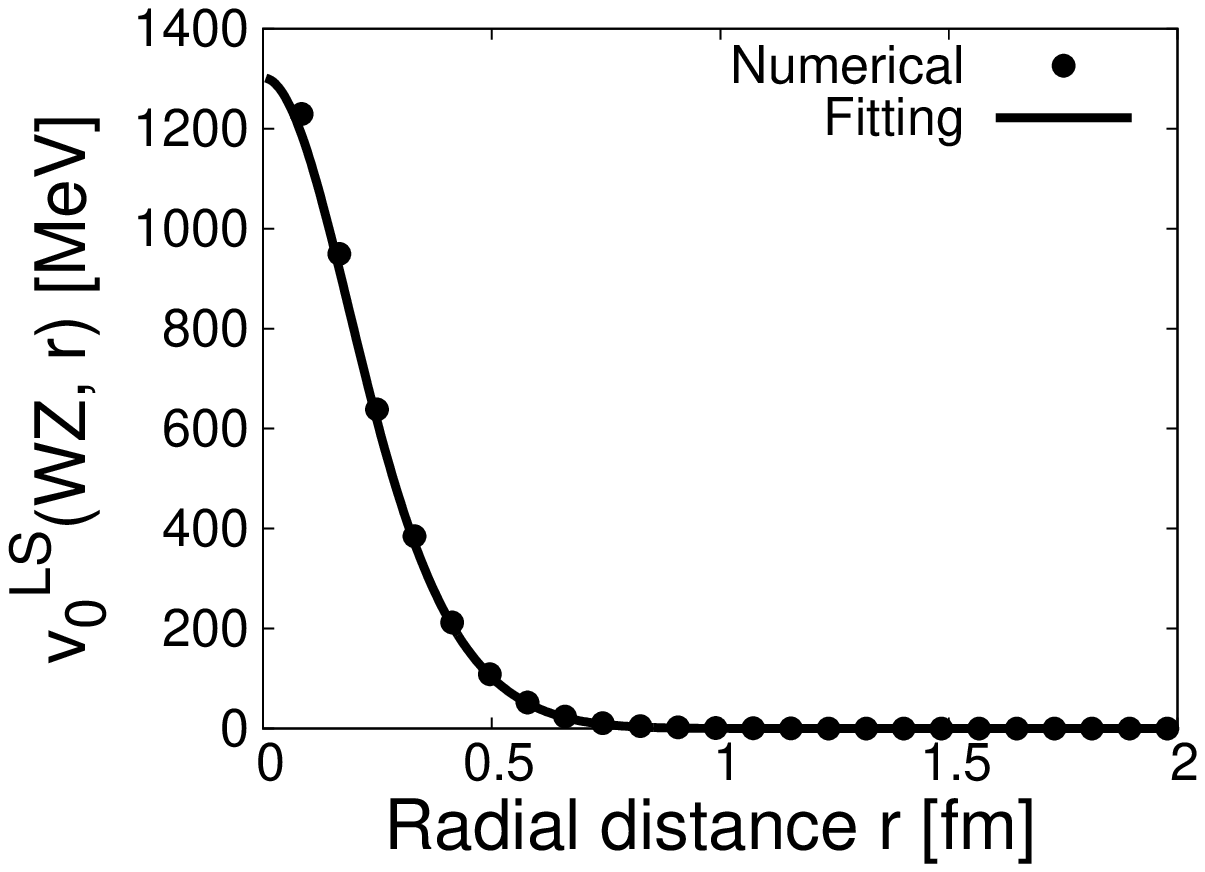}
	\end{center}
\end{minipage}
\caption{Comparisons between the numerically obtained and the Gaussian-fitted potentials 
	   for $u_{0}^{LS} \left(WZ, r \right)$~(left) and $v_{0}^{LS} \left(WZ, r \right)$~(right).}
\label{fig: comparison u_0^LS(WZ)}
\end{figure}
\begin{figure}[H]
\begin{minipage}{0.5\hsize}
	\begin{center}
		\includegraphics[width=8cm]{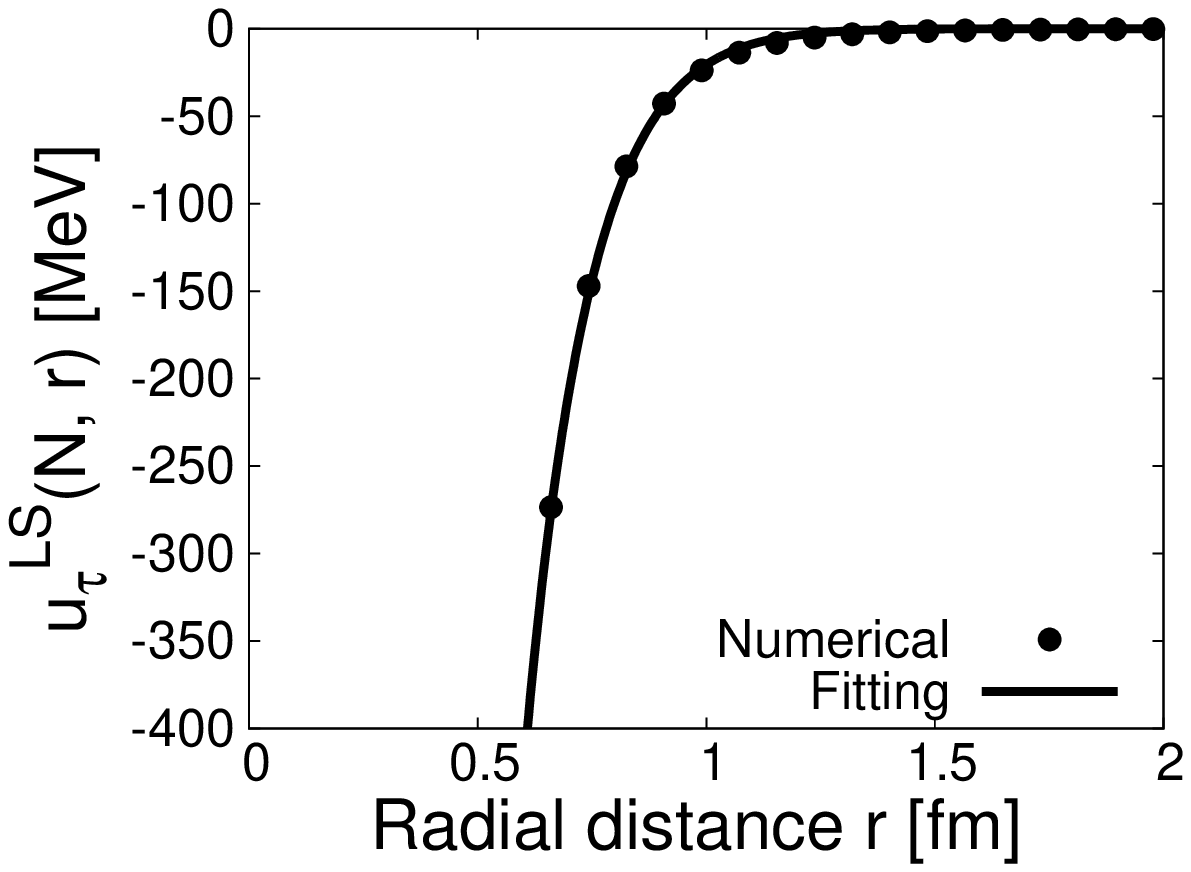}
	\end{center}
\end{minipage}%
\begin{minipage}{0.5\hsize}
 	\begin{center}
		\includegraphics[width=8cm]{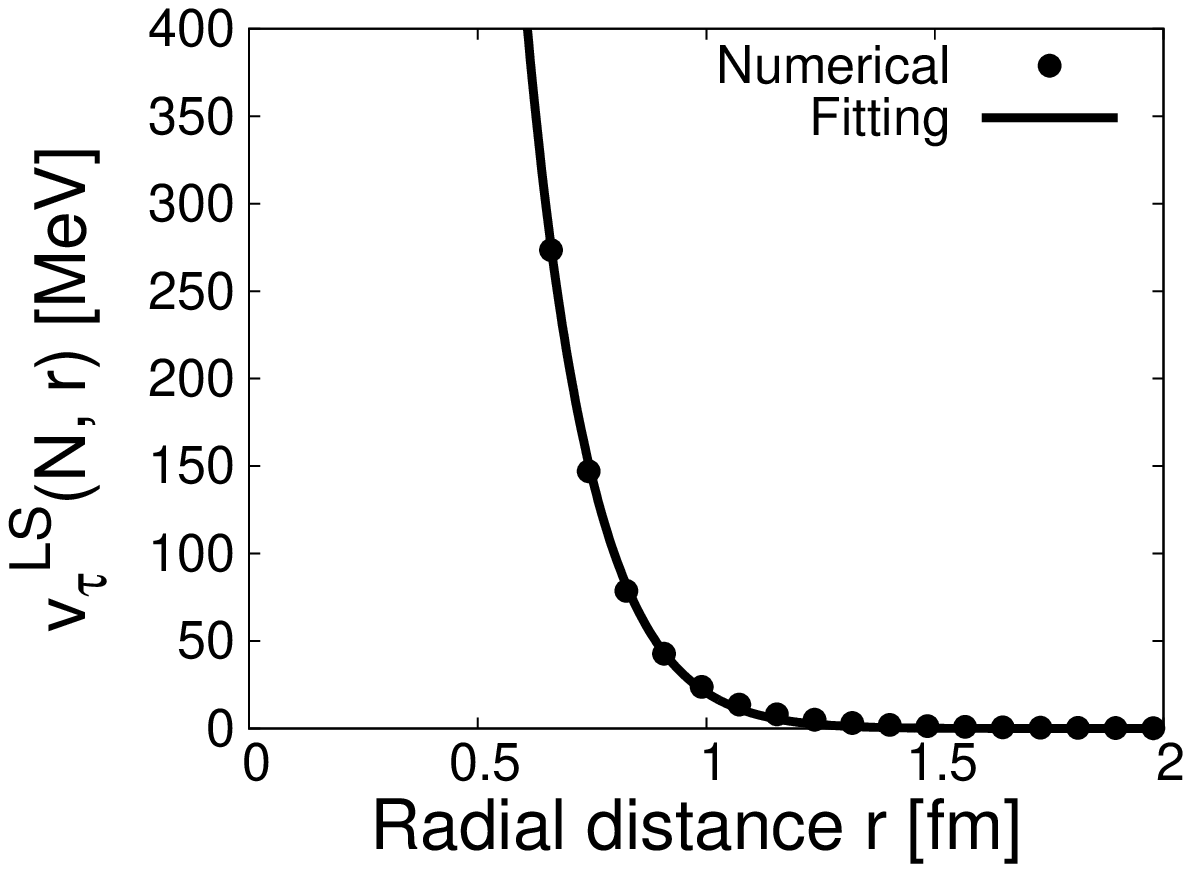}
	\end{center}
\end{minipage}
\caption{Comparisons between the numerically obtained and the Gaussian-fitted potentials 
	   for $u_{\tau}^{LS} \left(N, r \right)$~(left) and $v_{\tau}^{LS} \left(N, r \right)$~(right).}
\label{fig: comparison u_tau^LS}
\end{figure}
\begin{figure}[H]
\begin{minipage}{0.5\hsize}
	\begin{center}
		\includegraphics[width=8cm]{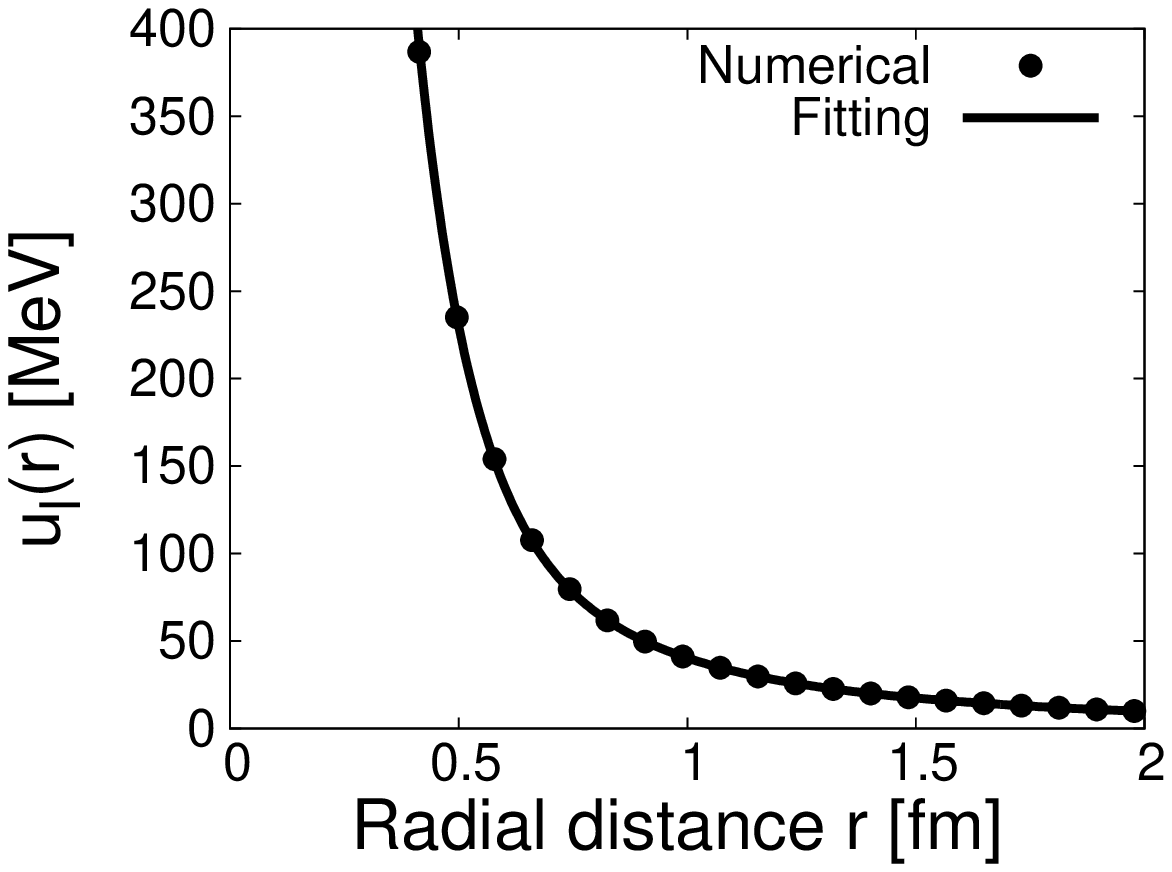}
	\end{center}
\end{minipage}%
\begin{minipage}{0.5\hsize}
 	\begin{center}
		\includegraphics[width=8cm]{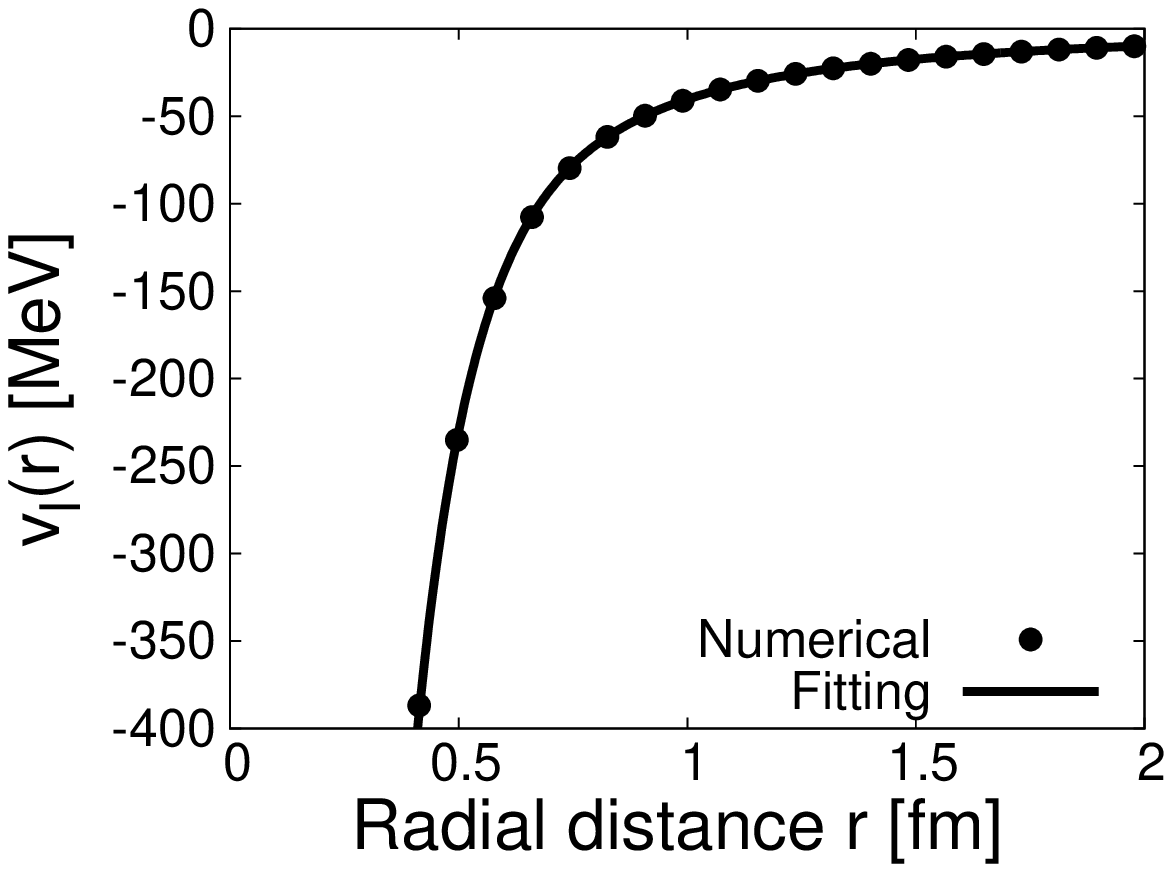}
	\end{center}
\end{minipage}
\caption{Comparisons between the numerically obtained and the Gaussian-fitted potentials 
	   for $u_l \left( r \right)$~(left) and $v_l \left( r \right)$~(right).}
\label{fig: comparison centrifugal term}
\end{figure}
From Figs.~\ref{fig: comparison u_0^c}~--~\ref{fig: comparison centrifugal term}, we can see that the fitting is done by the Gaussian forms in a good manner.
We find that four components, $U_0^c \left( WZ, r \right)$, $U_0^{LS} \left( N, r \right)$, $U_0^{LS} \left( WZ, r \right)$, and $U_{\tau}^{LS} \left( N, r \right)$, are fitted
by a single Gaussian function, $G_0 \left( r \right)$ or $G_{-2} \left( r \right)$, but with different ranges which are indicated by the superscript, while 
the others, $U_0^c \left( N, r \right)$, $U_{\tau}^c \left( N, r \right)$, and $U_{l} \left( r \right)$, are fitted with the different forms. 
We consider that the reason behind is that the former originates from a simple physical mechanism while the latte from complex one.


From now on, we make discussions for each component below.
\begin{itemize}
	\item Figure~\ref{fig: comparison u_0^c} \\
		For the s-wave channel, as shown in the upper left panel, we find that there is an attractive pocket whose depth is around 100 MeV in the middle range 
		and a repulsive core at a short distances in the energy independent potential, $u_0^c \left( N, r \right)$.
		Contrary, the energy dependent components, $v_0^c \left( N, r \right)$, behaves rather monotonically with an attraction as proportional to $1/r^2$.
		Turning to the p-wave potential as shown in the lower panel, energy independent component is attractive as proportional to $1/r^2$, 
		while the energy dependent one behaves similarly to the s-wave energy independent component, but with shorter range.  
		We can also see that the energy independent and dependent components behave in a quite different manner between the s- and p-wave channels.
		This is because of the nonlocal contributions of them.
		To see that, we first separate the potentials, $u_0^c \left( N, r \right)$ and $v_0^c \left( N, r \right)$, into the local and nonlocal contributions for the two channels,
		\bea
			u_0^c \left( N, r \right) &=& u_0^c \left( N, local, r \right) + u_0^c \left( N, non, r \right) \\
			v_0^c \left( N, r \right) &=& v_0^c \left( N, local, r \right) + v_0^c \left( N, non, r \right),
		\eea
		where the arguments $local$ and $non$ stand for the local and nonlocal contributions, respectively.
		Then, we have numerically calculated the nonlocal contributions for the s- and p-waves and shown the results in Fig.~\ref{fig: comparison nonlocal components} 
		where the s-wave components are plotted by solid line and the p-wave one by dashed line.  
		We can see that the nonlocal contributions are quite different between them.
		\begin{figure}[htb]
			\begin{minipage}{0.5\hsize}
			\begin{center}
				\includegraphics[width=8cm]{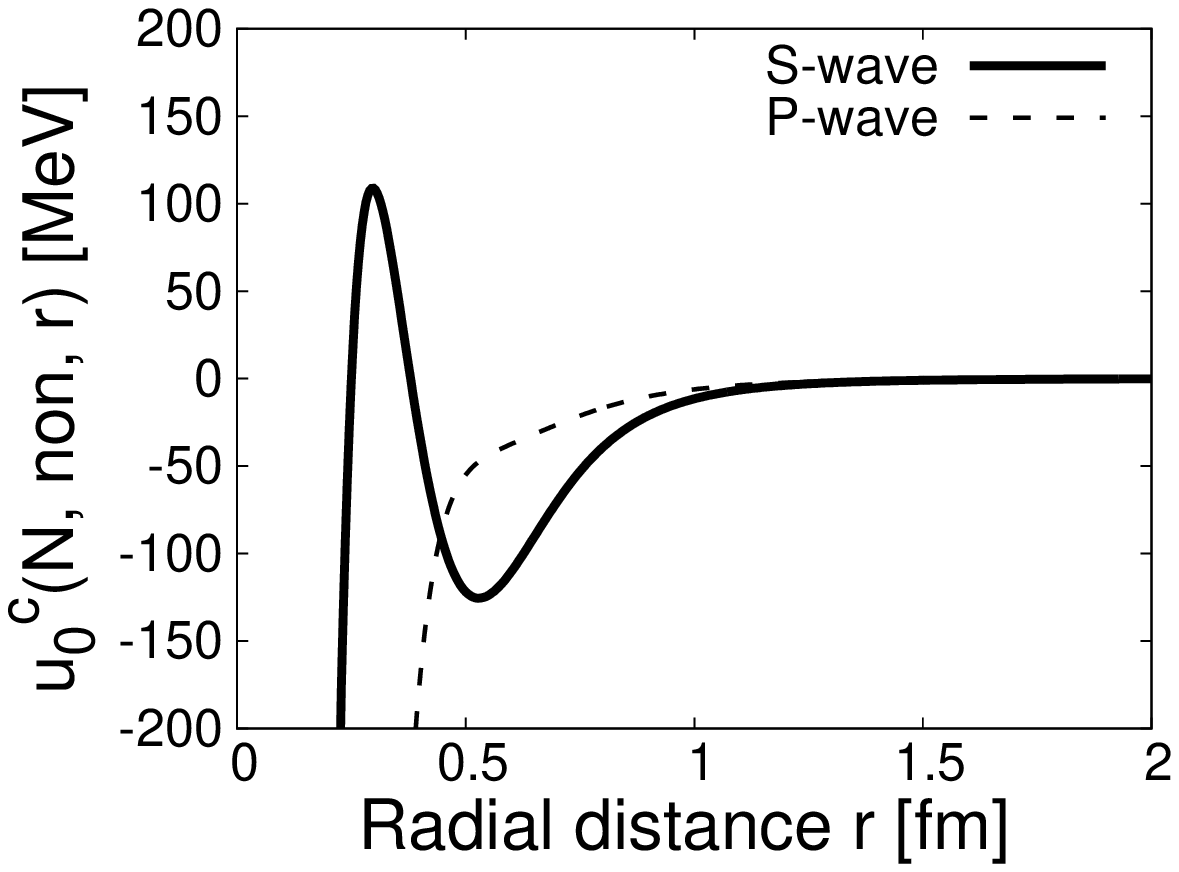}
			\end{center}
		\end{minipage}%
		\begin{minipage}{0.5\hsize}
		 	\begin{center}
				\includegraphics[width=8cm]{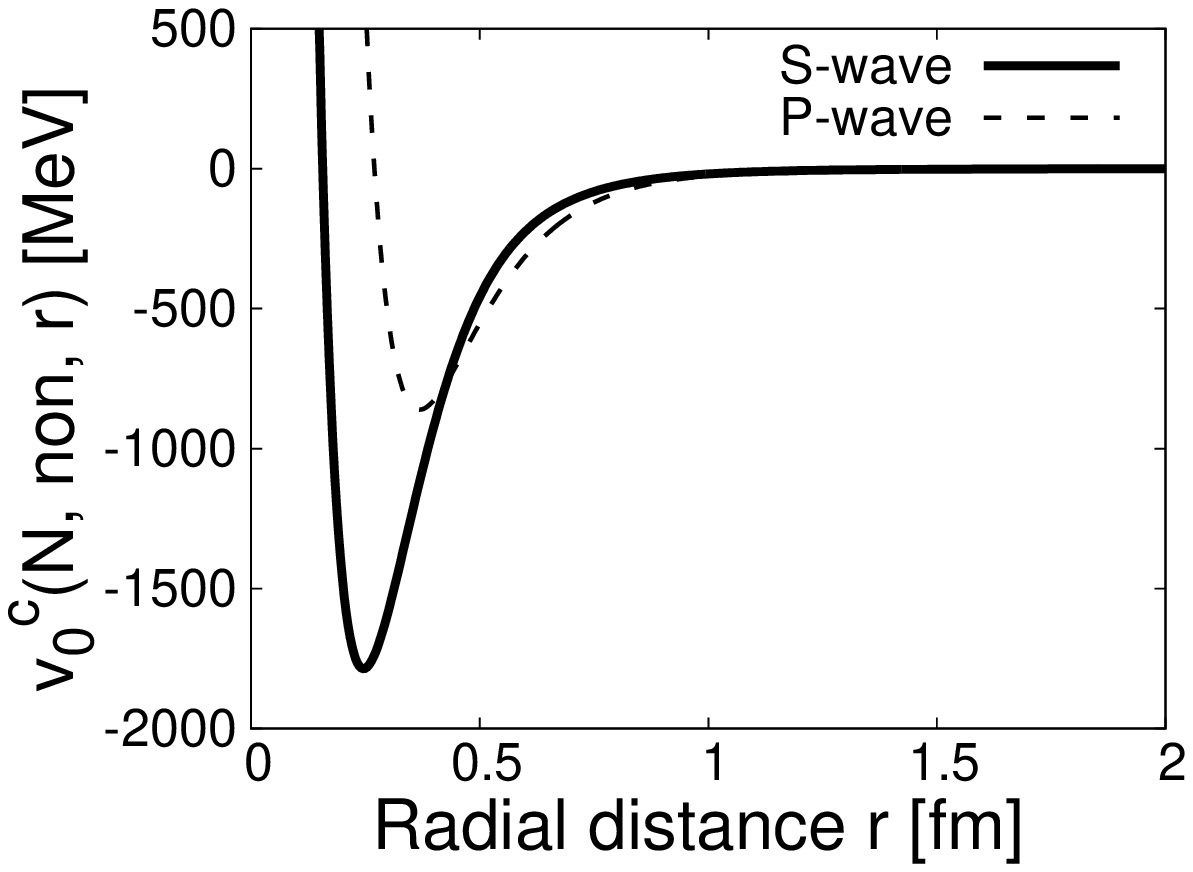}
			\end{center}
		\end{minipage}
		\caption{Nonlocal components of $u_{0}^{c} \left( N, r \right)$~(left) and $v_{0}^{c} \left( N, r \right)$~(right) for the s- and p-waves~(solid and dashed lines, respectively).}
		\label{fig: comparison nonlocal components}
		\end{figure}
		 		
	\item Figure~\ref{fig: comparison u_0^c(WZ)} \\
		We can see that the energy independent and dependent components behave in a similar way but their strengths are very much different.
		To see this, the contribution is expanded with respect to $\Delta_{E} = \varepsilon/2m_K$ as in Eq.~(\ref{eq: expanded potential}),
		\bea
			U_{0}^{c} \left( WZ, r \right) 
				&=& \displaystyle{\frac{1}{m_K + E}} \displaystyle{\frac{3}{\pi^2 F_{\pi}^2}} \displaystyle{\frac{\sin^2 F}{r^2}} F'  \left( E - \displaystyle{\frac{s^2}{\Lambda}} \right) \nonum \\
				&\propto& \displaystyle{\frac{1}{m_K + E}} \left( E - \displaystyle{\frac{s^2}{\Lambda}} \right) \nonum \\
				&\simeq& \left( \displaystyle{\frac{1}{2}} - \displaystyle{\frac{1}{2 m_{K}}} \displaystyle{\frac{s^2}{\Lambda}} \right)
							 + \left( \displaystyle{\frac{1}{2}} + \displaystyle{\frac{1}{2 m_{K}}} \displaystyle{\frac{s^2}{\Lambda}} \right)
							 	 \Delta_{E},
		\label{eq: expansion of WZ term}
		\eea
		where the explicit expressions of $U_{0}^{c} \left( WZ, r \right)$ is shown in Eq.~(\ref{eq:V0c}).
		In Eq.~(\ref{eq: expansion of WZ term}), $\Lambda$ is a moment of inertia of the SU(2) hedgehog soliton and we define $s = \sin \left( F\left( r \right) /2\right)$.
		In our definition for $u_{0}^{c} \left( WZ, r \right)$ and $v_{0}^{c}  \left( WZ, r \right)$,
		the first and second terms in Eq.~(\ref{eq: expansion of WZ term}) correspond to $u_{0}^{c} \left( WZ, r \right)$ and $v_{0}^{c}  \left( WZ, r \right)$, respectively.
		Therefore, we obtain the following relations,
		\bea
			u_{0}^{c} \left( WZ, r \right) \propto \displaystyle{\frac{1}{2}} - \displaystyle{\frac{1}{2 m_{K}}} \displaystyle{\frac{s^2}{\Lambda}}
		\label{eq: E-independent WZ term} 
		\eea
		and 
		\bea
			v_{0}^{c} \left( WZ, r \right) \propto \displaystyle{\frac{1}{2}} + \displaystyle{\frac{1}{2 m_{K}}} \displaystyle{\frac{s^2}{\Lambda}}. 
		\label{eq: E-dependent WZ term} 
		\eea 
		From these equations, we find that the difference of the energy independent and dependent terms of the Wess-Zumino term is proportional to ${s^2}/{\Lambda}$.
		This explains the difference in the strengths shown in Fig. 5.
		
	\item Figures~\ref{fig: comparison u_tau^c} and~\ref{fig: comparison u_0^LS} \\
		In these figures, it is shown that the energy independent and dependent components become exactly the same.
		This should have been expected from the analytic form of the potential as shown in Eqs.~(\ref{eq:Vtauc}) and~(\ref{eq:V0LS}), from which we can read explicitly, 
		\bea
			u \left( r \right) = v \left( r \right). 
		\eea
		 
	\item Figures ~\ref{fig: comparison u_0^LS(WZ)},~\ref{fig: comparison u_tau^LS}, and~\ref{fig: comparison centrifugal term} \\
		We can see that the strengths of the potentials are the same for each component but the sign is different between energy independent and dependent components.
		 We can easily verify it from their explicit forms shown in Appendix~\ref{appendix}.

\end{itemize}		 
\begin{figure}[htb]
\begin{minipage}{0.5\hsize}
	\begin{center}
		\includegraphics[width=8cm]{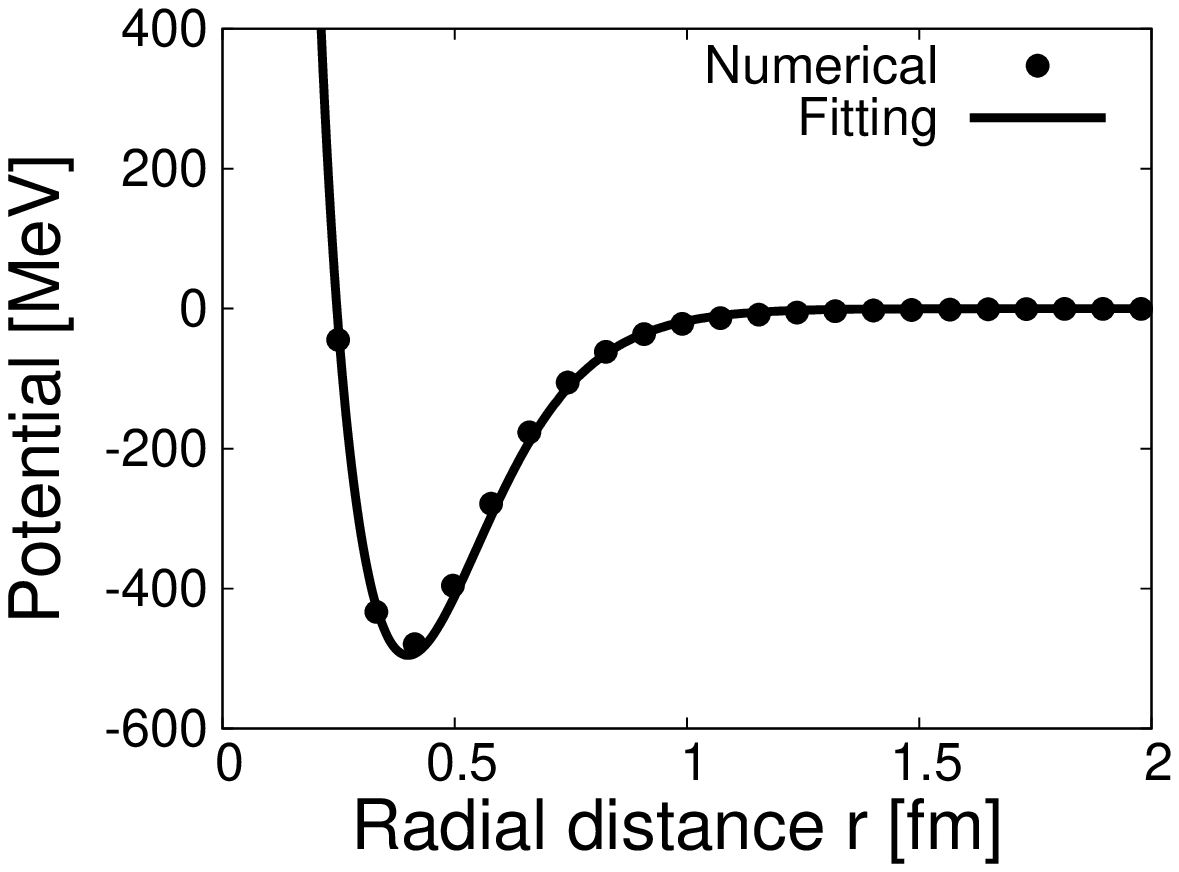}
	\end{center}
\end{minipage}%
\begin{minipage}{0.5\hsize}
 	\begin{center}
		\includegraphics[width=8cm]{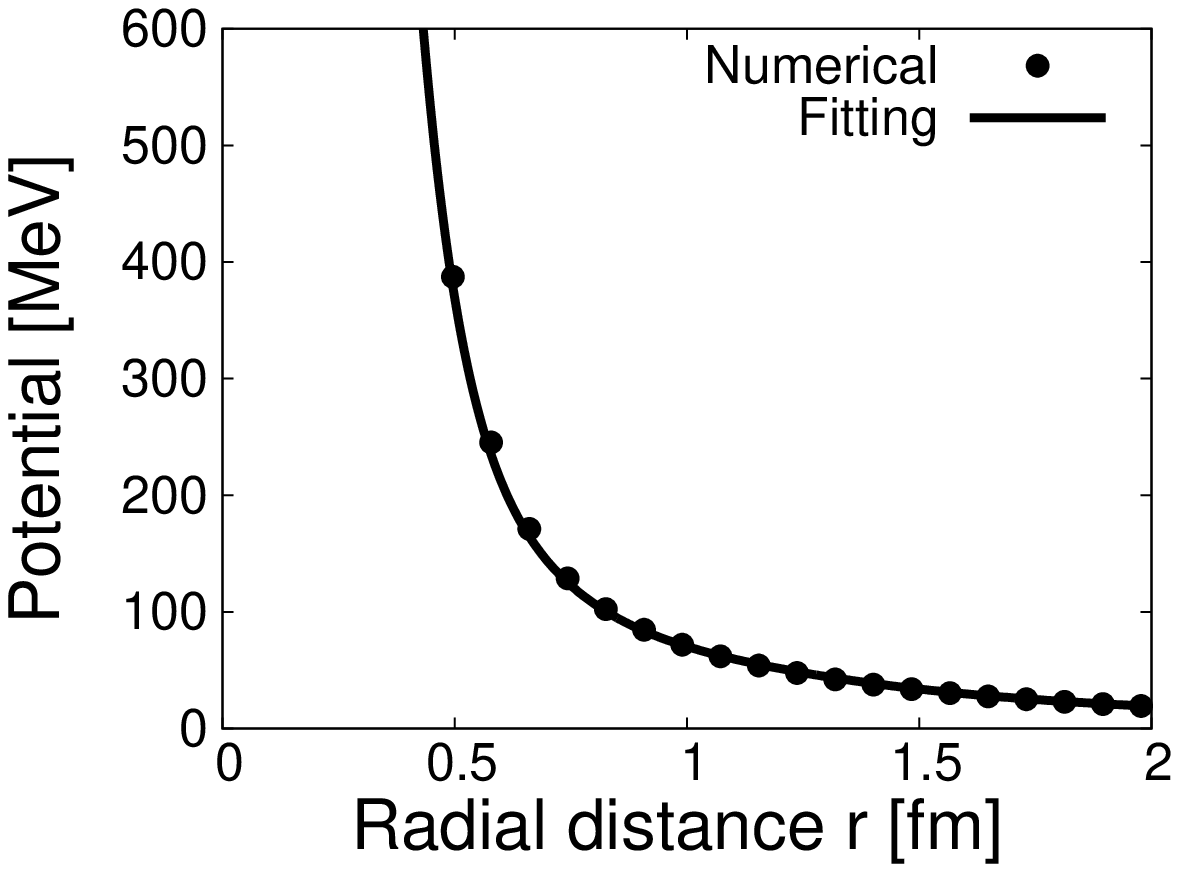}
	\end{center}
\end{minipage}
\caption{Total potentials for the $\bar{K}N \left( J^P =1/2^-, I = 0 \right)$ channel obtained from the bound state~(left) and
		 those for the $\bar{K}N \left( J^P = 3/2^+,I = 0 \right)$ channel from a scattering state~(right).}
\label{fig: comparison for set A}
\end{figure}

Finally, we show the total potentials which are numerically obtained and fitted by the Gaussian forms in Fig.~\ref{fig: comparison for set A}.
For the s-wave potential, we can see a repulsion at the short distances 
which comes from the isospin independent central term of the normal Skyrme Lagrangian, $U_0^c \left( N, r \right)$, as shown in Fig.~\ref{fig: comparison u_0^c}. 
In the middle range, we find an attractive pocket which may generate the bound state.
From Figs.~\ref{fig: comparison u_0^c}~--~\ref{fig: comparison u_tau^c}, this attractive pocket is dominantly mede by the attraction of the Wess-Zumino term. 

From Fig.~\ref{fig: comparison for set A}, we see that the behaviors of the potentials for the s-wave~($1/2^-$) and p-wave~($3/2^+$) are different; an attractive pocket vanishes for the p-wave. 
This is due to the strong repulsion of the LS and centrifugal components from the normal Lagrangian.

\begin{figure}[htb]
	\begin{center}
		\includegraphics[width=10cm]{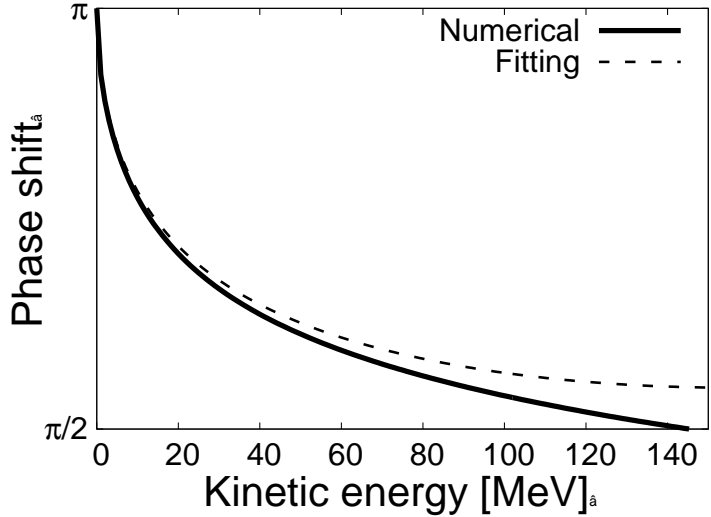}
	\end{center}
\caption{The phase shifts for the $\bar{K} N (J^P = 1/2^-, I = 0)$ channel obtained from the numerical and fitted potentials with the set A.}
\label{fig: comparison the phase shifts}
\end{figure}
So far we have seen that the fitting of the potential works well particularly for the local terms, 
while that for the nonlocal terms is not always the case as Fig.~\ref{fig: comparison nonlocal components} shows. 
The nonlocal terms induce also energy dependence.  
To see this point, we check how the phase shifts are reproduced by the fitted potential as functions of the kinetic energy.
In Fig.~\ref{fig: comparison the phase shifts}, we have compared the phase shifts clacurated by the numerically obtained and fitted potentials.
In the low energy region~$\left( \varepsilon \lesssim 50~\mathrm{MeV} \right)$ where we consider that our approach works well, the two phase shifts agree well.
Contrary, as the kinetic energy is getting larger, the difference of the phase shifts becomes larger, which is due to the nonlocal contributions.
Therefore, our fitted potential can be used for practical calculations for low energy kaon and nucleon systems.

\subsection{Scaling rules}
So far, we have performed the potential fitting for the parameter set A. 
In this subsection, we consider it for the sets B and C by using the scaling property of the Skyrmion.
In this way, various properties of the interaction will be better understood.  
First, we briefly review the scaling rule in the Skyrme model and then show the scaling rules for the fitting parameters. 
Finally, we compare the numerically obtained potential and fitted one from the parameter set A by scaling rules. 

\begin{figure}[htb]
	\begin{center}
		\includegraphics[width=10cm]{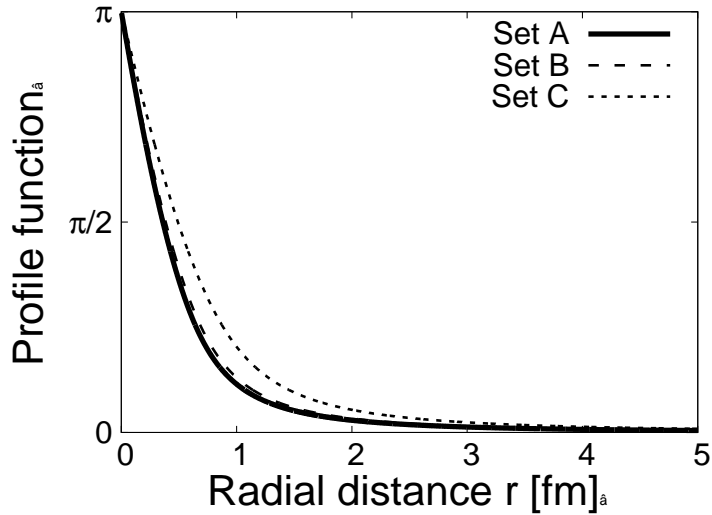}
	\end{center}
\caption{The profile functions $F \left( r \right)$ for the three parameter sets A, B, and C.}
\label{fig: profile functions}
\end{figure}

The Skyrme model of massless pion has one dimensionful parameter, $F_\pi$, and one coupling constant, $e$.  
These are scaled out by introducing the standard unit where length is expressed by 
\bea
	y = e F_{\pi} r.
\label{eq: scaling for distance}
\eea
By using this, soliton profiles for various $F_\pi$ and $e$ are related by a simple scale transformation to each other.
In Fig.~\ref{fig: profile functions}, we show the soliton profiles as functions of physical radial distance $r$ for the three parameter sets A, B, and C
which are obtained from the standard profile function with the scaling rule Eq.~(\ref{eq: scaling for distance}).
From Fig.~\ref{fig: profile functions}, we find that the profile function for the set C is most extended among the three parameter sets; soliton size is inversely proportional to $eF_{\pi}$.  

Having established the scaling rule for the soliton profile, let us investigate possible scaling rules for the kaon nucleon potential.  
First let us look at the relations among the parameter sets A, B, and C, and then we will discuss general cases.  

As expected from dimensional argument, it is shown that the range parameters for the parameter sets A and B, for instance, are related by
\bea
	R^{B} &=& \displaystyle{\frac{\alpha^A}{\alpha^B}} R^{A}
\label{eq: scaling rule 1} 
\eea	
for all components of the potential.
Here, we have defined $\alpha$ as $\alpha = e F_{\pi}$ and the superscripts $A$ and $B$ correspond to the parameter sets A and B, respectively, 
namely we take as follows, $\alpha^A = 4.67 \times 205$~MeV and $\alpha^B = 4.82 \times 186$~MeV.

Contrary, interaction strengths obey differently for different components.  
\begin{itemize}
	\item For the components which does not include $G_{-2}$ as a fitting function,
		 $U_{0}^{c} \left(WZ, r \right)$, $U_{\tau}^{c} \left(N, r \right)$, $U_{0}^{LS} \left(N, r \right)$, $U_{0}^{LS} \left(WZ, r \right)$, and $U_{l} \left( r \right)$, 
		 the strength parameters are scaled by the following rules, 
		\bea
			C_i^{B} = C_i^{A}, \ \ \ \left( i = -2, 0, 2 \right).
		\label{eq: scaling rule 2}
		\eea
	\item For the others~($U_{0}^{c} \left( N, r \right)$ and $U_{\tau}^{LS} \left( N, r \right)$), they obey the different rule as follows,
		\bea
			C_{i}^{B} 	&=&  \left( \displaystyle{\frac{\alpha^B}{\alpha^A}} \right)^2 C_{i}^{A}, \ \ \ \left( i = -2, 0, 2 \right). 
		\label{eq: scaling rule 3}
		\eea
\end{itemize}
 
\begin{figure}[htb]
\begin{minipage}{0.5\hsize}
	\begin{center}
		\includegraphics[width=8cm]{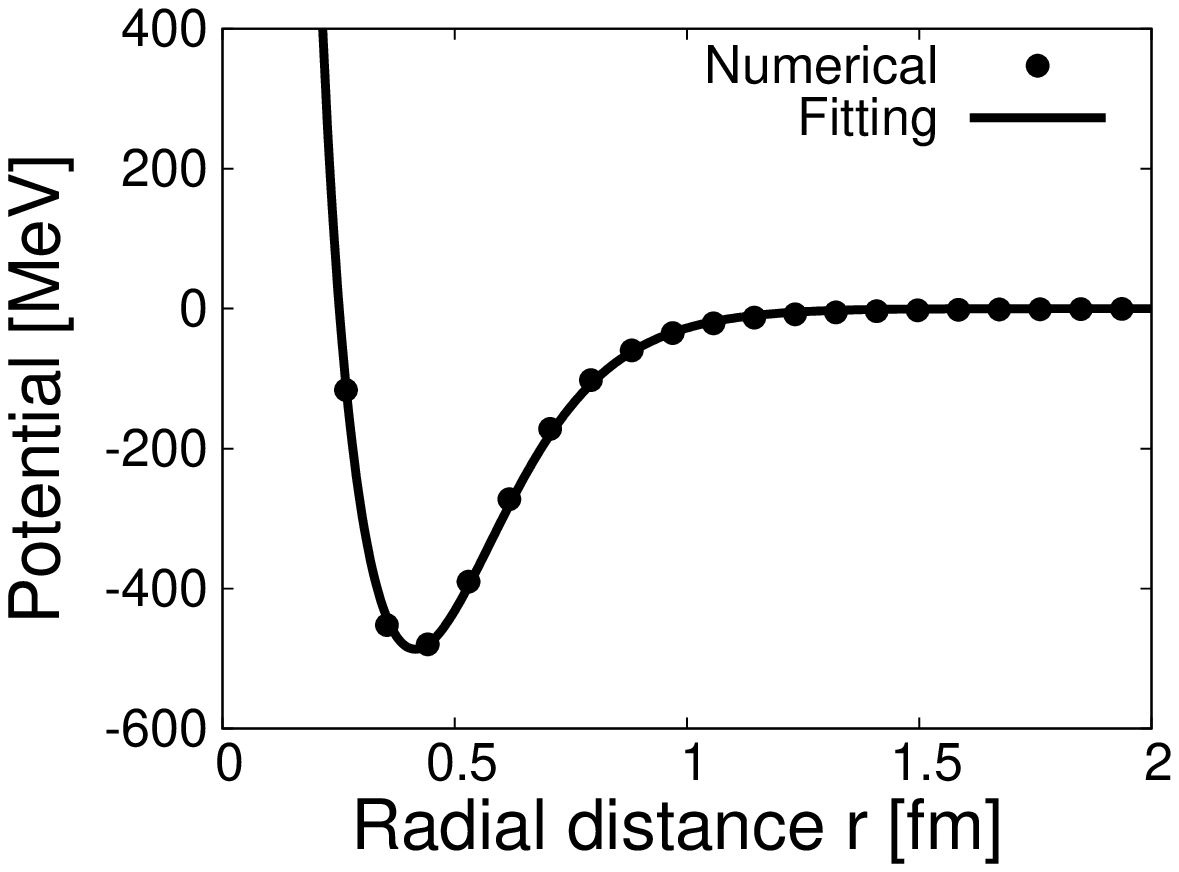}
	\end{center}
\end{minipage}%
\begin{minipage}{0.5\hsize}
 	\begin{center}
		\includegraphics[width=8cm]{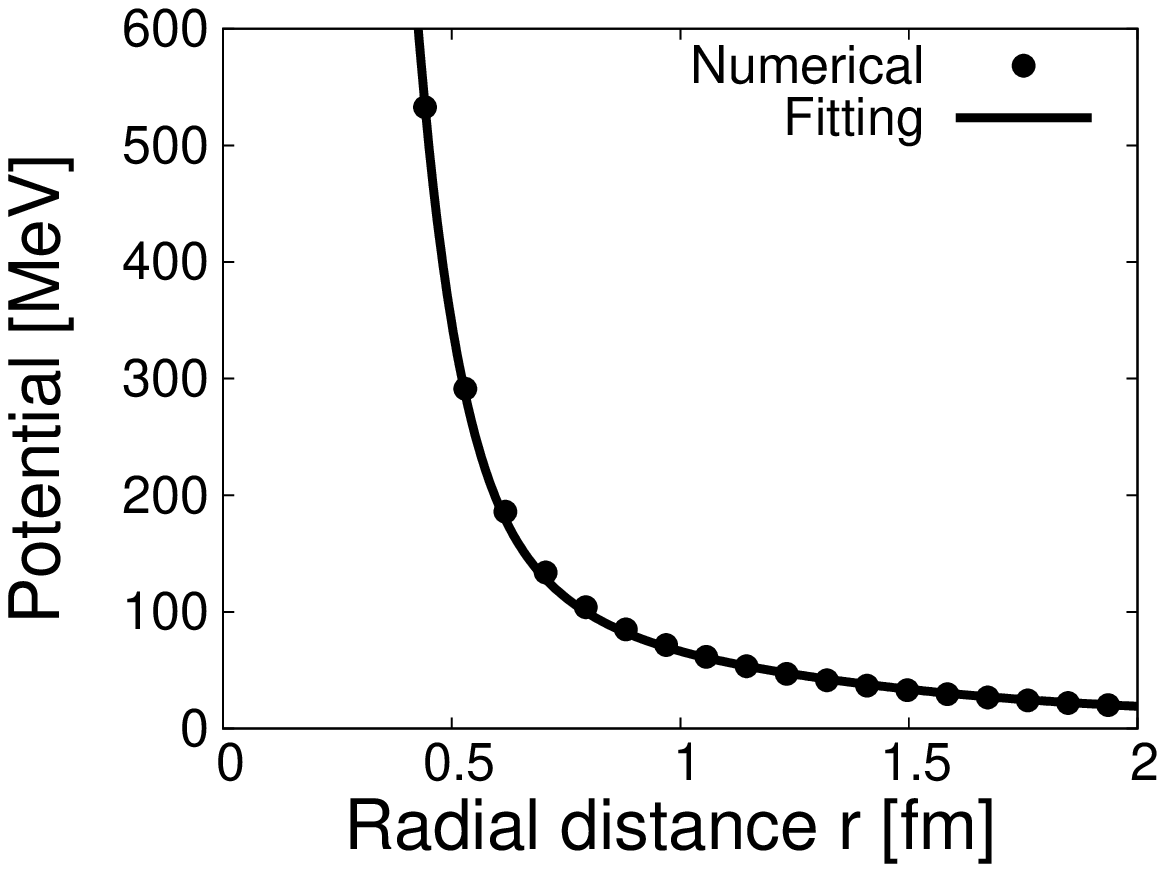}
	\end{center}
\end{minipage}
\caption{Total potentials for the $\bar{K}N \left( J^P =1/2^-, I = 0 \right)$ channel from the bound state~(left) and
		 the $\bar{K}N \left( J^P = 3/2^+,I = 0 \right)$ channel from a scattering state~(right) for the parameter set B
		 which are derived from the potentials for the set A by the scaling rules Eqs.~(\ref{eq: scaling rule 1})~--~(\ref{eq: scaling rule 3}).
		 The scattering energies are -32.2 and 27 MeV for s- and p-waves, respectively.}
\label{fig: comparison for set B}
\end{figure}

\begin{figure}[htb]
\begin{minipage}{0.5\hsize}
	\begin{center}
		\includegraphics[width=8cm]{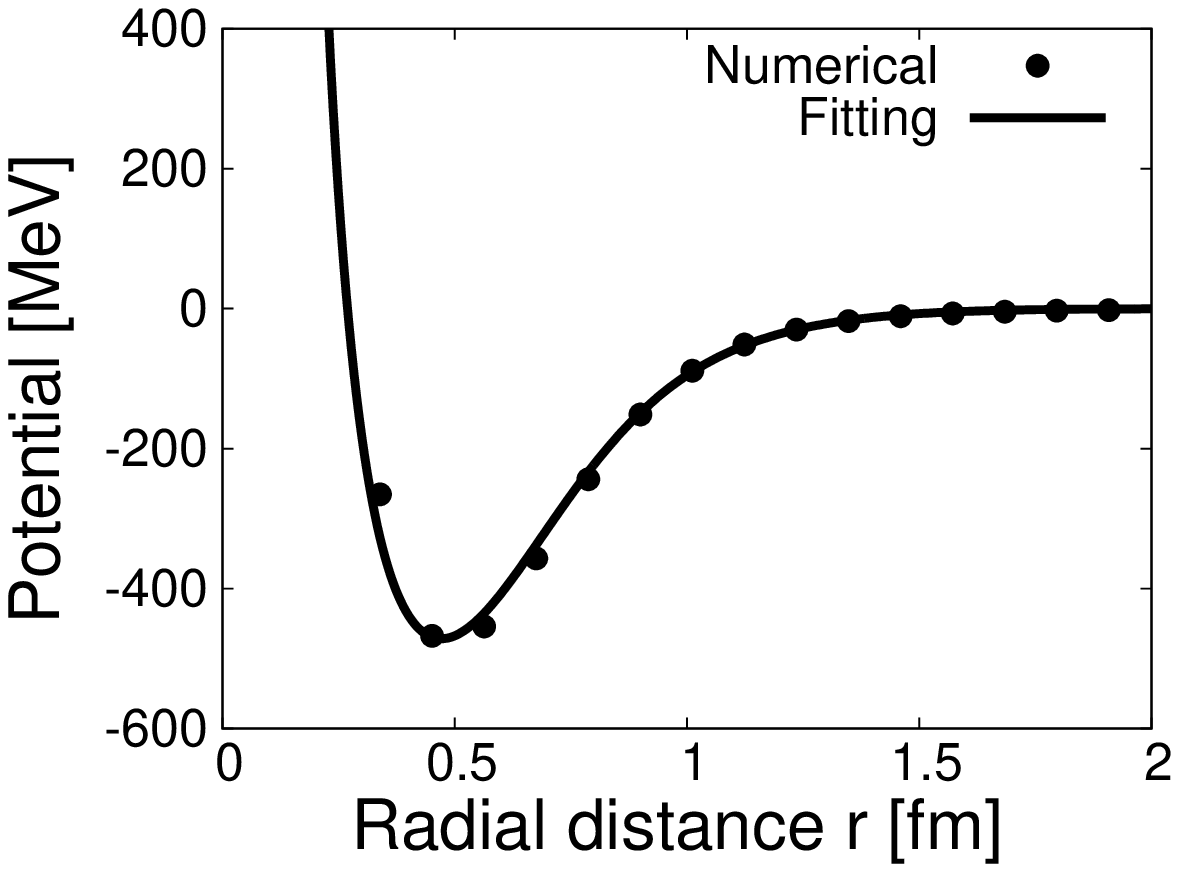}
	\end{center}
\end{minipage}%
\begin{minipage}{0.5\hsize}
 	\begin{center}
		\includegraphics[width=8cm]{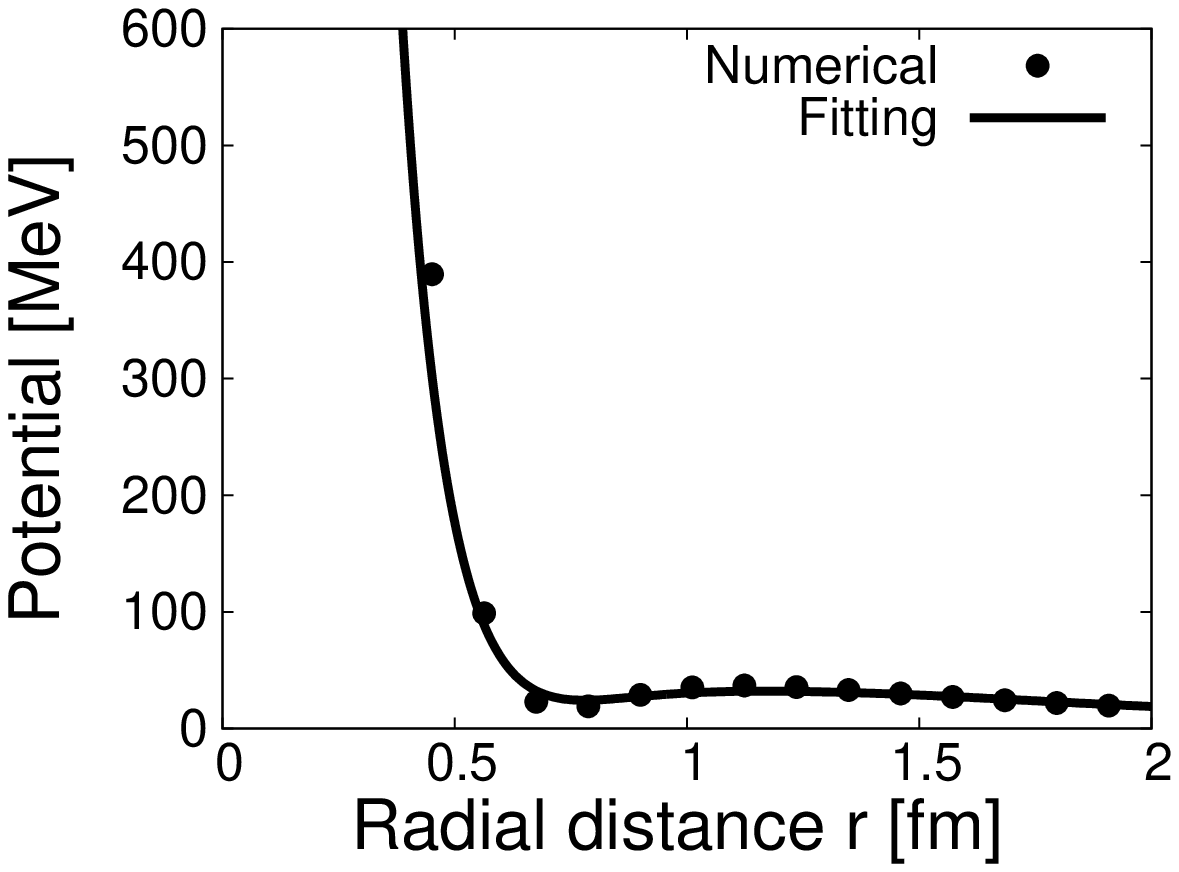}
	\end{center}
\end{minipage}
\caption{Total potentials for the $\bar{K}N \left( J^P =1/2^-, I = 0 \right)$ channel from the bound state~(left) and
		 the $\bar{K}N \left( J^P = 3/2^+,I = 0 \right)$ channel from a scattering state~(right) for the parameter set C 
		 which are derived from the potentials for the set A by the scaling rules Eqs.~(\ref{eq: scaling rule 1})~--~(\ref{eq: scaling rule 3}).
		 The scattering energies are -81.3 and 27 MeV for s- and p-waves, respectively.}
\label{fig: comparison for set C}
\end{figure}

In Fig.~\ref{fig: comparison for set B} and~\ref{fig: comparison for set C}, we have shown the potentials for the same channel as in Fig.~\ref{fig: comparison for set A} for  the parameter sets B and C.
The s-wave potentials are calculated at the binding energies of the corresponding parameter set, namely, $\varepsilon = -32.2$~MeV for the set B and -81.3~MeV for the set C. 
The p-wave potential is calculated at the common scattering energy $\varepsilon = 27$~MeV. 
From Figs.~\ref{fig: comparison for set B} and~\ref{fig: comparison for set C}, 
we find that the potentials for the parameter set A is scaled into the sets B and C with the scaling rules Eqs.~(\ref{eq: scaling rule 1})~--~(\ref{eq: scaling rule 3}) in a good manner.

\begin{figure}[H]
\begin{minipage}{0.5\hsize}
	\begin{center}
		\includegraphics[width=8cm]{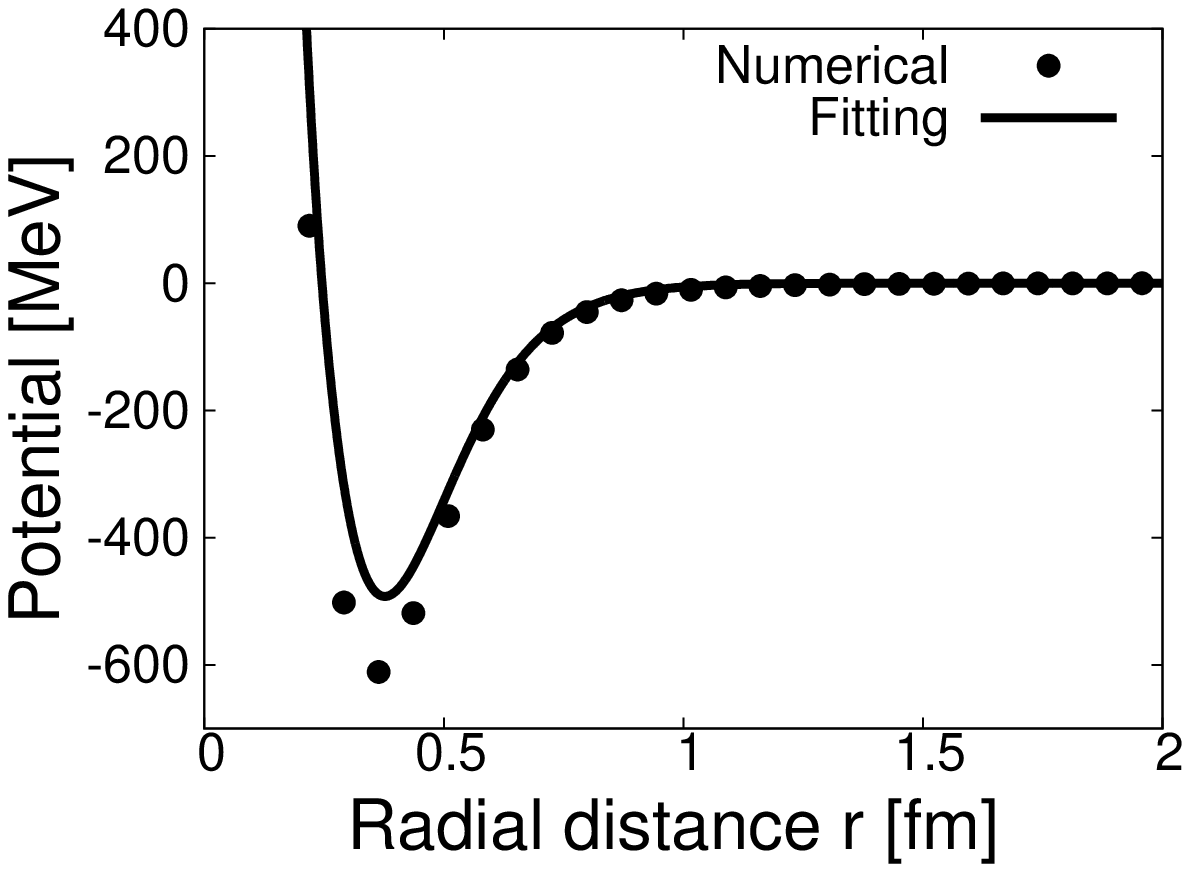}
	\end{center}
\end{minipage}%
\begin{minipage}{0.5\hsize}
 	\begin{center}
		\includegraphics[width=8cm]{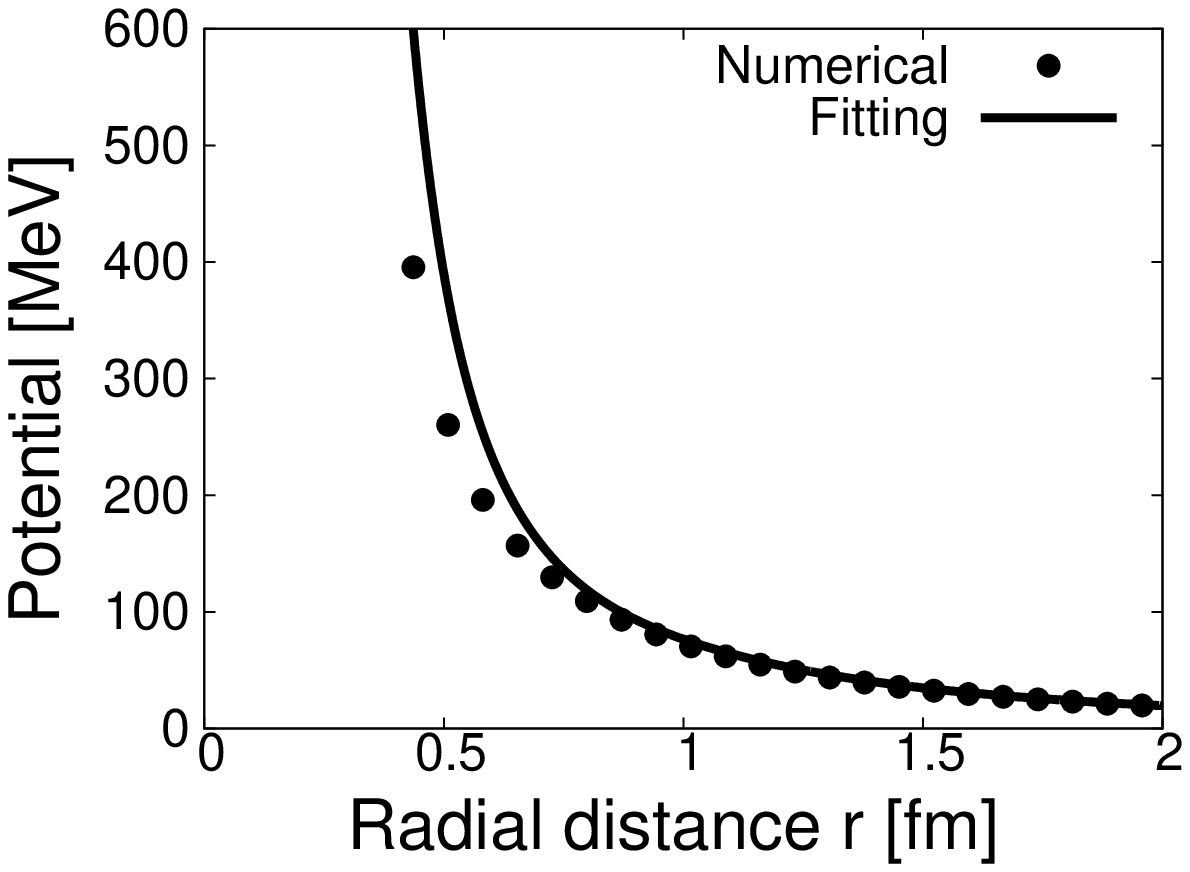}
	\end{center}
\end{minipage}
\caption{Potentials for the $\bar{K}N \left( J^P =1/2^-, I = 0 \right)$ channel from the bound state~(left) and
		 the $\bar{K}N \left( J^P = 3/2^+,I = 0 \right)$ channel from a scattering state~(right) for the parameter set D 
		 which are derived from the potentials for the set A by the scaling rules Eqs.~(\ref{eq: scaling rule 1})~--~(\ref{eq: scaling rule 3}).
		 The scattering energies are -21.0 and 27 MeV for s- and p-waves, respectively.}
\label{fig: comparison for set D}
\end{figure}

Finally, we consider general parameter sets.
To do that, let us first observe that the potential contains terms with different $1/N_c$ behaviors,
the one originates from the soliton profile~(leading order term)
and the one from the rotation~(higher order term).
The former is factored out by 1 in the standard unit, while the latter by $e^3F_{\pi} \sim 1/N_c$ which is inversely proportional to the moment of inertia.  
Because of this, the scaling rules for different parameter sets of $F_{\pi}$ and $e$ differ for these two terms of different $1/N_c$ orders.
For the case of set A, B and C, because the parameters are chosen to preserve the value $e^3F_{\pi}$ unchanged, 
we have obtained a simple scaling rule as dictated by Eqs.~(\ref{eq: scaling rule 1})~--~(\ref{eq: scaling rule 3}).
In general this is no longer the case and we have to consider the scaling rules for the leading and higher order terms, separately.
To see how the simple scaling rule holds generally, we introduce a new parameter set D which is taken as $F_{\pi} = 186$~MeV and $e = 5.85$.
In this parameter set, we set the pion decay constant at the experimental value,
while the Skyrme parameter at the $\rho \pi \pi$-coupling constant, $g_{\rho \pi \pi}$, determined from the KSRF relation~\cite{KSRF 1, KSRF 2},
\bea
	m_{\rho}^2 = \displaystyle{\frac{g_{\rho \pi \pi}^2 F_{\pi}^2}{2}},
\label{eq: KSRF relation}
\eea
where $m_{\rho} = 770$~MeV which is the mass of the $\rho$-meson.
We show the potential calculated by the set D and those expected by the scalings Eqs.~(\ref{eq: scaling rule 1})~--~(\ref{eq: scaling rule 3}) in Fig.~\ref{fig: comparison for set D}.
The binding energy of the $\bar{K} N$ bound state is 21.0~MeV for the set D and the scattering energy is for 27~MeV for the p-wave.
There is some deviation between the two, which is not, however, very large.
To conclude this subsection, the potential obeys a simple scaling rule as long as the moment of inertia is unchanged.

\section{Summary}
In this paper, we have discussed the kaon-nucleon scattering states and the kaon-nucleon potentials by a modified bound state approach in the Skyrme model~\cite{ezoe-hosaka}.
In our approach, the potential contains terms of different orders of $1/N_c$ due to the change of projection and variation.
In the limit $N_c \to \infty$, this violates the $1/N_c$ expansion series, but we think it reasonable for physical systems of weakly interacting kaon and nucleon, which may generate molecular like states.  

First, we have investigated the phase shifts for the kaon-nucleon scattering with lower partial waves of the kaon.
The obtained phase shifts indicate that the potential is attractive for the $\bar{K}N$ channel and repulsive for the $KN$ one.
Then, we have evaluated the scattering length for the $\bar{K}N$ scattering state but it is larger than experimental results and other theoretical calculations~\cite{scattering length 1, scattering length 2, scattering length 3, scattering length 4, scattering length 5}.

Second, to make further discussions for the potential, we have classified the $\bar{K}N$ potential into the seven components according to their natures with and without energy dependence.
Then, we have fitted them by the Gaussian type functions.
As a results, we have found that all the components can be fitted by the Gaussian type functions.
Actually, we have verified that the binding energies of the $\bar{K} N$ bound state and the phase shifts derived from the fitted potential and those from the numerically obtained original ones agree well.  

While fitting the potential, we have also investigated the scaling rules associated with the soliton profile function, when changing the model parameters $F_{\pi}$ and $e$.   
We have found that various components of the potential contain terms of different order of $1/N_c$, which obey different scaling rules, separately.
However, by keeping the moment of inertia in the higher order terms unchanged when $F_{\pi}$ and $e$ are varied, these separate scaling rules reduce to a simple rule for each component.  

For further studies, it would be necessary to take into account the finite mass effect of the pion.  
Furthermore, coupling of $\bar{K} N$ and $\pi \Sigma$ is important.  
Improvements by considering these aspects should be performed for more quantitative discussions of kaon-nucleon systems including $\Lambda(1405)$.  

\begin{acknowledgments}
We thank Noriyoshi Ishii for useful discussions.
This work is supported in part by the Grant-in-Aid  for Science Research (C) JP26400273.
\end{acknowledgments}

\appendix
\section{The equation of motion and interactions}
\label{appendix}
In this appendix, we show the explicit expressions of various terms in the equation of motion Eq.~(\ref{eq: equation of motion}),
\bea
	- \displaystyle{\frac{1}{r^2}} \displaystyle{\frac{d}{dr}} \left( r^2 h\left( r \right) \displaystyle{\frac{dk_l^{\alpha} \left( r \right)}{dr}}\right)
		- E^2 f \left( r \right) k_l^{\alpha} \left( r \right)
		+ \left( m_K^2 + V \left( r \right) \right) k_l^{\alpha} \left( r \right) = 0,
\label{eq: equation of motion for kaon}
\eea
where
\bea
	h(r) = 1 + \displaystyle{\frac{1}{\left( e F_{\pi} \right)^2}} \displaystyle{\frac{2}{r^2}} \sin^2 F, 
\label{eq:h}
\eea
\bea
	f(r) = 1 + \displaystyle{\frac{1}{\left( e F_{\pi} \right)^2}} \left( \displaystyle{\frac{2}{r^2}} \sin^2 F + F'^2 \right),
\label{eq:f}
\eea
\bea
	V \left( r \right) = V_0^c \left( r \right) + V_{\tau}^c \left( r \right) I_{KN} + V_0^{LS} \left( r \right) J_{KN} + V_{\tau}^{LS} \left( r \right) J_{KN} I_{KN}.
\label{eq:potential}
\eea
In Eq.~(\ref{eq:potential}), we define $I_{KN}$ and $J_{KN}$ as follows
\bea
	I_{KN} = \bm{I}^K \cdot \bm{I}^N, \ \ \ \
	J_{KN} = \bm{L}^K \cdot \bm{J}^N,
\label{eq:spin and isospin products}
\eea
where the nucleon spin and isospin operators, $\bm{J}^N$ and $\bm{I}^N$, are given by~\cite{zahed}
\bea
	\bm{J}^N = i \Lambda \mathrm{tr} \left[ \bm{\tau} \dot{A}^{\dag} \left( t \right) A \left( t \right) \right], \ \ \ \ 
	\bm{I}^N = i \Lambda \mathrm{tr} \left[ \bm{\tau} \dot{A} \left( t \right) A^{\dag} \left( t \right) \right].
\label{eq: spin and isospin operator}
\eea

In Eq.~(\ref{eq: spin and isospin operator}), $\dot{A} \left( t \right)$ is the time derivative of an isospin rotation matrix, $A \left( t \right)$, 
$\bm{\tau}$ are the $2 \times 2$ Pauli matrices, and $\Lambda$ is the soliton moment of inertia given by~\cite{ANW}
\bea
	\Lambda = \displaystyle{\frac{2 \pi}{3}} F_{\pi}^2 \int dr \ r^2 \sin^2 F \left[ 1 + \displaystyle{\frac{4}{\left( e F_{\pi} \right)^2}} \left( F'^2 + \displaystyle{\frac{\sin^2 F}{r^2}} \right) \right].
\label{eq: moment of inertia}
\eea
The isospin and angular momentum operators of the kaon, $\bm{I}^K$ and $\bm{L}^K$, are given by
\bea
	\bm{I}^K= \displaystyle{\frac{\bm{\tau}}{2}}, \ \ \ \ 
	\bm{L}^K = \bm{r} \times \bm{p}^K,
\eea
where $\bm{p}^K$ is the momentum of the kaon.

Finally, the explicit forms of each term in the interaction, $V \left( r \right)$, are given by the following equations.
\bea
	V_0^c \left( r \right)
		&=& - \displaystyle{\frac{1}{4}} \left( 2 \frac{\sin^2 F}{r^2} + (F')^2 \right) 
			+ 2 \displaystyle{\frac{s^4}{r^2}}
			+ \left[ 1 + \displaystyle{\frac{1}{\left( e F_{\pi} \right)^2}} \left( F'^2 + \displaystyle{\frac{\sin^2 F}{r^2}} \right)  \right] \displaystyle{\frac{l \left( l + 1 \right)}{r^2}}  \nonum \\
		&& - \displaystyle{\frac{1}{\left( e F_{\pi} \right)^2}}  \left[ 2 \displaystyle{\frac{\sin^2 F}{r^2}} \left( \displaystyle{\frac{\sin^2 F}{r^2}} + 2 (F')^2 \right) 
												   - 2 \displaystyle{\frac{s^4}{r^2}} \left( F'^2 + \displaystyle{\frac{\sin^2 F}{r^2}} \right) \right] \nonum \\
		&& + \displaystyle{\frac{1}{\left( e F_{\pi} \right)^2}} \displaystyle{\frac{6}{r^2}} \left[ \displaystyle{\frac{s^4 \sin^2 F}{r^2}} + \displaystyle{\frac{d}{dr}} \left\{ s^2 \sin F F' \right\} \right] \nonum \\
		&& + \displaystyle{\frac{2 E}{\Lambda}} s^2 \left[ 1 + \displaystyle{\frac{1}{\left( e F_{\pi} \right)^2}} \left( F'^2 + \displaystyle{\frac{5}{r^2}} \sin^2 F \right) \right] \nonum \\
		&& + \displaystyle{\frac{3}{\left( e F_{\pi} \right)^2}} \displaystyle{\frac{1}{r^2}} \displaystyle{\frac{d}{dr}} \left[ r^2 \left( \displaystyle{\frac{E F' \sin F }{\Lambda}}  \right)  \right] 
			\pm \displaystyle{\frac{3}{\pi^2 F_{\pi}^2}} \displaystyle{\frac{\sin^2 F}{r^2}} F'  \left( E - \displaystyle{\frac{s^2}{\Lambda}} \right),
\label{eq:V0c}
\eea
\bea
	V_{\tau}^c \left( r \right)
		&=& \displaystyle{\frac{8 E}{3 \Lambda}} s^2 \left[ 1 + \displaystyle{\frac{1}{\left( e F_{\pi} \right)^2}}  \left( F'^2 + \displaystyle{\frac{4}{r^2}} \sin^2 F \right) \right]
			+ \displaystyle{\frac{4}{\left( e F_{\pi} \right)^2}} \displaystyle{\frac{1}{r^2}} \displaystyle{\frac{d}{dr}} \left[ r^2 \left( \displaystyle{\frac{ E F' \sin F}{\Lambda}} \right)  \right], \nonum \\
\label{eq:Vtauc}
\eea
\bea
	V_0^{LS} \left( r \right)
		&=& \displaystyle{\frac{1}{\left( e F_{\pi} \right)^2}} \displaystyle{\frac{2E \sin^2 F}{\Lambda r^2}} 
			\pm \displaystyle{\frac{3}{F_{\pi}^2 \pi^2}} \displaystyle{\frac{\sin^2 F}{\Lambda r^2}} F',
\label{eq:V0LS}
\eea
and
\bea
	V_{\tau}^{LS} \left( r \right)
		&=& - \left[ 1 + \displaystyle{\frac{1}{\left( e F_{\pi} \right)^2}}  \left( F'^2 + 4 \displaystyle{\frac{\sin^2 F}{r^2}} \right) \right] \displaystyle{\frac{16 s^2}{3 r^2}} 
			- \displaystyle{\frac{1}{\left( e F_{\pi} \right)^2}}\displaystyle{\frac{8}{r^2}} \left[ \displaystyle{\frac{d}{dr}}\left( \sin F F'\right) \right],
\label{eq:VtauLS}
\eea
where
\bea
	s = \sin \left( F \left( r \right) / 2 \right), 
\eea
and
\bea
	F' = dF \left( r \right)/dr.
\eea
The last terms of Eq.~(\ref{eq:V0c}) and Eq.~(\ref{eq:V0LS}) are derived from the Wess-Zumino term, which is attractive for the $\bar{K}N$ potential and repulsive for the $KN$ potential.

\section{Fitting parameters}
\label{app: fitting parameters}
In this appendix, we show the fitting parameters discussed in Sec.~\ref{sec: numerical fitting}
\begin{table}[htb]
\begin{minipage}{0.5\hsize}
	\begin{center}
		\begin{tabular}{| c || c | c | c |} \hline
		    							& $G_{-2} \left( r \right)$	&  $G_{0} \left( r \right)$ 	& $G_{2} \left( r \right)$	\\ \hline 
			Range [fm]				& 0.165   				& 0.254 				& 0.368     			\\              
                    	$u_0^c \left( N, r \right)$ [MeV]	& 3320             			& 2903				& -579				\\ 
                    	$v_0^c \left( N, r \right)$ [MeV]	& -2244				& -6343				& -499			\\ \hline
		\end{tabular}
	\end{center}
\end{minipage}%
\begin{minipage}{0.5\hsize}
	\begin{center}
		\begin{tabular}{| c || c | c | c |} \hline
		    							& $G_{-2} \left( r \right)$	&  $G_{0} \left( r \right)$ 	& $G_{2} \left( r \right)$	\\ \hline 
			Range [fm]				& 0.298   				& 0.292				& 0.300     			\\              
                    	$u_0^c \left( N, r \right)$ [MeV]	& -3161             		& 2185				& -504 				\\ 
                    	$v_0^c \left( N, r \right)$ [MeV]	& 2965				& -3249				& -1995				\\ \hline
		\end{tabular}
	\end{center}
\end{minipage}
\caption{Fitting parameters for $u_0^c \left( N, r \right)$ and $v_0^c \left( N, r \right)$ for the s-wave~(left) and p-wave~(right).}
\label{tab: normal term isospin independent central term}
\end{table}
\begin{table}[H]
	\begin{center}
		\begin{tabular}{| c || c | c | c |} \hline
		    								&  $G_{0}^{(1)} \left( r \right)$ 	& $G_{0}^{(2)} \left( r \right)$	\\ \hline 
			Range [fm]					& 0.264   					& 0.378 					\\              
                    	$u_{0}^c \left( WZ, r \right)$ [MeV]	& -677      		       			& -1207					\\ 
                    	$v_{0}^c \left( WZ, r \right)$ [MeV]	& -3449					& -985					\\ \hline
		\end{tabular}
	\end{center}
\caption{Fitting parameters for $u_{0}^c \left( WZ, r \right)$ and $v_{\tau}^c \left( WZ, r \right)$.}
\label{tab: WZ term isospin dependent central term}
\end{table}
\begin{table}[H]
	\begin{center}
		\begin{tabular}{| c || c | c | c |} \hline
		    								&  $G_{0} \left( r \right)$ 	& $G_{2} \left( r \right)$	\\ \hline 
			Range [fm]					& 0.248   				& 0.491 				\\              
                    	$u_{\tau}^c \left( N, r \right)$ [MeV]	& 401	             		& 291				\\ 
                    	$v_{\tau}^c \left( N, r \right)$ [MeV]	& 401				& 291				\\ \hline
		\end{tabular}
	\end{center}
\caption{Fitting parameters for $u_{\tau}^c \left( N, r \right)$ and $v_{\tau}^c \left( N, r \right)$.}
\label{tab: normal term isospin dependent central term}
\end{table}
\begin{table}[H]
	\begin{center}
		\begin{tabular}{| c || c | c | c |} \hline
		    								&  $G_{0}^{(1)} \left( r \right)$ 	& $G_{0}^{(2)} \left( r \right)$	\\ \hline 
			Range [fm]					& 0.281 					& 0.452   					\\              
                    	$u_0^{LS} \left( N, r \right)$ [MeV]	& 127					& 78						\\ 
                    	$v_0^{LS} \left( N, r \right)$ [MeV]	& 125					& 78						\\ \hline
		\end{tabular}
	\end{center}
\caption{Fitting parameters for $u_0^{LS} \left( N, r \right)$ and $v_0^{LS} \left( N, r \right)$.}
\label{tab: normal term isospin independent LS term}
\end{table}
\begin{table}[H]
	\begin{center}
		\begin{tabular}{| c || c | c | c |} \hline
		    									&  $G_{0}^{(1)} \left( r \right)$ 	& $G_{0}^{(2)} \left( r \right)$	\\ \hline 
			Range [fm]						& 0.228					& 0.353   					\\              
                    	$u_{0}^{LS} \left( WZ, r \right)$ [MeV]	& -574					& -728	             			\\ 
                    	$v_{0}^{LS} \left( WZ, r \right)$ [MeV]	& 574					& 728	             			\\ \hline
		\end{tabular}
	\end{center}
\caption{Fitting parameters for $u_{0}^{LS} \left( WZ, r \right)$ and $v_{0}^{LS} \left( WZ, r \right)$.}
\label{tab: WZ term isospin dependent LS term}
\end{table}
\begin{table}[H]
	\begin{center}
		\begin{tabular}{| c || c | c | c |} \hline
		    									&  $G_{-2}^{(1)} \left( r \right)$ 	& $G_{-2}^{(2)} \left( r \right)$	\\ \hline 
			Range [fm]						& 0.245					& 0.566   	 				\\              
                    	$u_{\tau}^{LS} \left( N, r \right)$ [MeV]	& -7930					& -1465 					\\ 
                    	$v_{\tau}^{LS} \left( N, r \right)$ [MeV]	& 7930					& 1465					\\ \hline
		\end{tabular}
	\end{center}
\caption{Fitting parameters for $u_{\tau}^{LS} \left( N, r \right)$ and $v_{\tau}^{LS} \left( N, r \right)$.}
\label{tab: normal term isospin dependent LS term}
\end{table}	
\begin{table}[H]
	\begin{center}
		\begin{tabular}{| c || c | c | c |} \hline
		    										&  $G_{0}^{(1)} \left( r \right)$ 	& $G_{0}^{(2)} \left( r \right)$	\\ \hline 
			Range [fm]							& 0.404  					& 0.700					\\              
                    	$u_l \left( r \right)$ [MeV$^2 \cdot$fm$^2$ ]	& 66226           				& 6074					\\ 
                    	$v_l \left( r \right)$ [MeV$^2 \cdot$fm$^2$ ]	& -66228					&  -6074					\\ \hline
		\end{tabular}
	\end{center}
\caption{Fitting parameters for the centrifugal force.}
\label{tab: centrifugal force}
\end{table}


\end{document}